# A REVIEW OF MATHEMATICAL AND COMPUTATIONAL METHODS IN CANCER DYNAMICS

Abicumaran Uthamacumaran[1], Hector Zenil[2,3,4,5]

[1]Concordia University, Department of Physics (Alumni), Montreal, QC, Canada
[2]The Alan Turing Institute, British Library, London, NW1 2DB, U.K.
[3]Oxford Immune Algorithmics, Reading, RG30 1EU, U.K.
[4]Algorithmic Dynamics Lab, Karolinska Institute, Stockholm, 171 77, Sweden
[5]Algorithmic Nature Group, LABORES, Paris, 76006, France

Cancers are complex adaptive diseases regulated by the nonlinear feedback systems between genetic instabilities, environmental signals, cellular protein flows, and gene regulatory networks. Understanding the cybernetics of cancer requires the integration of information dynamics across multidimensional spatiotemporal scales, including genetic, transcriptional, metabolic, proteomic, epigenetic, and multi-cellular networks. However, the time-series analysis of these complex networks remains vastly absent in cancer research. With longitudinal screening and time-series analysis of cellular dynamics, universally observed causal patterns pertaining to dynamical systems, may self-organize in the signaling or gene expression state-space of cancer triggering processes. A class of these patterns, strange attractors, may be mathematical biomarkers of cancer progression. The emergence of intracellular chaos and chaotic cell population dynamics remains a new paradigm in systems medicine. As such, chaotic and complex dynamics are discussed as mathematical hallmarks of cancer cell fate dynamics herein. Given the assumption that time-resolved single-cell datasets are made available, a survey of interdisciplinary tools and algorithms from complexity theory, are hereby reviewed to investigate critical phenomena and chaotic dynamics in cancer ecosystems. To conclude, the perspective cultivates an intuition for computational systems oncology in terms of nonlinear dynamics, information theory, inverse problems, and complexity.  We highlight the limitations we see in the area of statistical machine learning but the opportunity at combining it with the symbolic computational power offered by the mathematical tools explored.

## 1. INTRODUCTION

Cancer is the second leading cause of disease-related death globally, and a tremendous burden to progressive medicine. Deciphering the minimal set of interactions in the complex multiscale networks driving cancer gene expression and signaling remains an intractable problem due to its collective emergent behaviors, including phenotypic (epigenetic) plasticity, intra-tumoral heterogeneity, therapy resistance, and cancer stemness (Hanahan, 2022). Cancer is essentially a genetic disease and given the vast amounts of scientific works on the genetic and chromosomal instabilities/mutations driving tumorigenesis and cancer evolutionary dynamics, we shall not directly explore these conventional realms herein. Rather the review will focus on the cell fate dynamics emerging from these genetic/chromosomal instabilities, and hallmarks of cancer progression/adaptivity. Further, molecular heterogeneity is observed across many scales of cancer cybernetics, necessitating multiomics and multimodal profiling methods to dissect cancer ecosystems (Zahir et al., 2020). In this regards, systems biology and computational medicine have paved many powerful tools in single-cell analyses including network theory, data science, statistical machine learning, and multivariate information theoretics. However, current approaches in network medicine rely on static (single time-point) visualizations of spatial cancer cell-signaling and gene expression profiles. These predominant snapshot approaches are fundamental limiting factors in the advancement of precision oncology since they are causal agnostic, i.e., they remove the notion of *time* (*dynamics*) from cancer datasets. Instead, pseudotemporal ordering techniques are used to infer gene expression patterns and cell fate trajectories in cancer processes on dimensionality-reduced pattern spaces. Although these static patterns may provide us insights into statistical correlations in complex cancer processes, statistical *correlation does not imply causation*. The lack of time-series measurements in single-cell multi-omics (e.g., gene expression dynamics, protein oscillations, histone marks spreading, etc.) and cell population fluctuations (i.e., ecological dynamics), in patient-derived tumor and liquid biopsies, remains a central roadblock in reconstructing cancer networks as complex *dynamical systems*.

Let us suppose we do have time-sequential measurements to infer signaling and gene expression dynamics in cancer networks by the methods suggested above. What kinds of patterns emerge in time-series which cannot be inferred from the currently predominant snapshot approaches? How do we detect these causal patterns in cancer dynamics? The review was written precisely to address these questions and provide a general intuition for nonlinear dynamics, chaos, and complexity in cancer research. The central dogma in systems biology remains that gene expression dynamics are *stochastic processes*. However, unlike stochastic systems, *deterministic chaos*, although difficult to distinguish from randomness, has a defined causal pattern in state-space (Gleick, 2008). Chaotic systems exhibit an underlying (multi)fractal topology, defined as strange attractor(s). Unlike randomness, these causal patterns allow the short-term predictions of the chaotic system's time-evolution (up to some Lyapunov time) and their global state-space trajectory (orbit). Chaotic systems also exhibit patterns of *emergent behaviors*, i.e., collective patterns and structures which are unpredictable from the individual components (Gleick, 2008).

The primary theme of the paper is, if chaotic behavior or complex dynamics plays an important role within cancer dynamics, how do we characterize chaos/complexity and distinguish it from randomness? The review comprises of a detailed discussion of tools from complex systems science and nonlinear dynamics (dynamical systems theory) to decode multiscale processes and behavioral patterns in cancer cellular dynamics. Various detecting tools, measures and algorithms exist which remain under-explored in (computational) systems medicine. Complex multiscale/multiomic networks driving cancer emergence/progression are discussed across different scales of cancer processes, from gene regulatory networks to epigenetic stemness/plasticity networks. An intuition for cellular oscillations (both intracellular flows and population dynamics) and causality inference in cancer cell fate dynamics are presented under the lens of complex systems. The remaining bulk of the paper is devoted to a step-by-step blueprint of algorithms and tools to capture multiscale chaotic dynamics (if they exist) within cancer cell signaling and cellular processes. Dynamical systems theory is a relatively new framework to most cancer researchers further stressing the lack of time-series cancer datasets. As such, some codes for a selected set detection tools/algorithms are provided in the appendix. A summary table of some inference methods discussed are also provided. Furthermore, traditionally, signaling refers to protein-protein interactions or signal transduction pathways in cell communication networks. Examples of such signaling includes physiological cybernetics such as psychoneuroendocrine control, immune-inflammatory pathways, neurotransmitter dynamics, extracellular vesicles-mediated communication networks, and other receptor-ligand regulatory feedback loops/signaling cascades. However, we may simply refer to various scales of cancer cybernetics, including cancer-immune population dynamics, protein density fluctuations, epigenetic patterns/chromatin modifications, metabolomics, and gene expression (transcriptional) dynamics, to name a few, simply as *signaling* herein in reference signal (information) processing in control systems (cybernetics). Due to the limited space allocated to the comprehensive review, additional information such as biological insights into epigenetic complexity, some fine-details of the mathematical treatments/methods, and prospective techniques for the acquisition of time-resolved multicell data are provided In the Appendix.

## 2. COMPLEX DYNAMICS

*Complexity theory* is an interdisciplinary paradigm in systems science merging nonlinear dynamics, statistical mechanics, information theory, and computational physics. It deals with the study of *whole systems* which exhibit emergent behavioral patterns, often due to their multi-scale *nonlinear interactions*, and multi-nested feedback loops. Therefore, *complex systems* are (in general) nonlinear feedback systems (including computational systems) with many interacting parts which give rise to collective behaviors (i.e., emergence) (Nicolis and Rouvas-Nicolis, 2007). Complex systems or their signature, emergence, may be best defined by the non-reductionistic Aristotelian dictum *the whole is more than the sum of its parts* (Wolfram, 1988). Complexity theory is thus the quantitative study of collective processes, patterns, and behaviors in complex systems. *Chaotic systems* are at the heart of biological/physiological complex systems and warrant our deepest attention.

The universality of chaotic dynamics in physiological control systems has been well-established. Based on the findings of the Jacob-Monod model of lac operon regulation, Goodwin, a student of Waddington, first derived an oscillatory model of gene regulatory networks to describe negative feedback loop oscillators as seen in a wide range of biological processes including circadian rhythms, enzymatic processes, developmental biology, and cell cycle dynamics (Goodwin, 1965). Later, the Mackey-Glass equations, a set of first-order nonlinear delay-differential equations, demonstrated that complex dynamics ranging from limit-cycle oscillations to chaotic attractors can emerge in respiratory and hematopoietic diseases (Mackey and Glass, 1977). The most prominent examples of chaotic dynamics are found within cardiac oscillations (Glass et al., 1987; Goldberger et al., 1990; Skinner et al., 1990). The works of Winfree and Kuramoto further extended the study of biological oscillators in physiological control processes/rhythms (Winfree, 1967, Kuramoto, 1984). Chaotic oscillations have also been well-studied in glycolytic oscillations and cellular calcium fluxes. For instance, tumor glycolytic oscillations have been experimentally suggested to confer adaptive cellular behaviors such as therapy resistance in tumor ecosystems. In this model system, Pomuceno-Ordunez et al. (2019) investigated the effects of pulses and periodic glucose deprivation in a kinetic model of HeLa cells glycolysis. A system of ordinary differential equations were obtained from the model to quantify the glycolytic oscillations. Various measures were used to assess the complex dynamics including **stability analysis** of the steady state, stroboscopic analysis, and Lempel-Ziv index. The study concluded that periodic glucose pulses can lead to an increase in the energy charge, while periodic glucose deprivation of the tumor ecosystem prevented the increase in the complexity of glycolytic oscillations and caused a decrease in the cellular energy charge of tumor cells (Pomuceno-Orduñez et al., 2019). However, it should be emphasized that complex tumor ecosystems such as GBM exhibit adaptive heterogeneity and phenotypic plasticity amidst a diverse range of transcriptional and metabolic cellular states. While some cellular states may be glycolytic phenotypes, others may favor oxidative phosphorylation, and others are inclined towards other metabolic programs.

Further, the detection of chaotic oscillations in other cellular rhythms such as the circadian clock remain experimentally dormant. Only mathematical models and numerical simulations have by far shown the emergence of chaotic behavior at the level of clock protein oscillations in simple model systems like Drosophila (Gonze et al., 2000). The detection of intracellular chaotic oscillations in proteins and genetic networks, is recently emerging as a *paradigm shift* in complexity science (Jensen et al., 2012; Heltberg et al., 2019). Most biological systems at varying length and time scales, including networks of genes, proteins, and populations of cells, behave like coupled nonlinear oscillators (Strogatz and Stewart, 1993; Strogatz, 2004). In principle, chaotic oscillations can arise in these biological oscillators (Strogatz and Stewart, 1993). In the context of cancer networks, there are many timescales and interconnected regulatory feedback loops manifesting time-delays in their oscillatory dynamics. The time-delays may give birth to signaling cascades and a symphony of complex dynamics, including chaotic oscillations (Strogatz, 2004; Strogatz, 2015). For instance, calcium oscillations within cells are in the timescales of seconds, whereas protein oscillations such as transcription factor oscillations, the cell cycle, and circadian rhythms span from hours to days. This is the key insight cancer researchers should be aware of, that oscillatory dynamics occur in all length and time scales, including networks of gene expression (transcriptional dynamics), protein signals, multicellular networks, and ecological/population dynamics (e.g., tumor-immune predator-prey systems). When these oscillations become aperiodic and irregular, they may either be stochastic (random) or chaotic. However, unlike randomness, chaotic flows have an underlying causal pattern, a structure, in state-space to which their irregular trajectories are confined to (Strogatz, 2015). The methods for detecting different behavioral regimes in the experimental time-traces of cancer signals depend on the timescales of the oscillations and the multi-nestedness (interactions) of the complex network patterns or dynamics they form. This, the resolution of the time-series datasets acquired must also be considered in chaos/complex dynamics discovery.

The time-series signal, whether it be the oscillation of a single protein, a protein concentration density during cellular patterning, or gene expression dynamics, of cancer cells, can be represented as a state-vector $X(t)$. For instance, the gene expression matrix acquired from a single-cell RNA-Seq experiment is a state vector at a given time point t. The state-space, also known as phase-space, determines the set of all possible values of the signal's state-vector (Strogatz, 2015). Any state of the dynamical system at a moment frozen in time can be represented as a point in phase space. All the information about its position and velocity is contained in the coordinates of that fixed-point. As the system evolves, the point would trace a trajectory in phase space (Strogatz, 2015). In the context of the given example, the state-space would describe the entire range of possibilities in the oscillator(s), or all the possible gene-gene network configurations described by the count matrix. The state-space reconstruction of the signals' time-traces (trajectories) exhibit a set of universal patterns called *attractors*. Attractors are self-organized causal structures governing the fate of a dynamical system in state-space. They are finite regions bound to state-space to which the trajectories of the dynamical system are confined to or pulled towards (i.e., attracted to) (Note: the opposite flow analog also exists, repellors, regions in state space from which the system is pushed away from) (Strogatz, 2015). The detection of attractors provide a route to reduce the combinatorially vast state-space of all network configurations conferring cancer states towards a finite set of values. However, only fixed-point attractors (equilibrium points), the simplest of attractors, are analytically solvable.

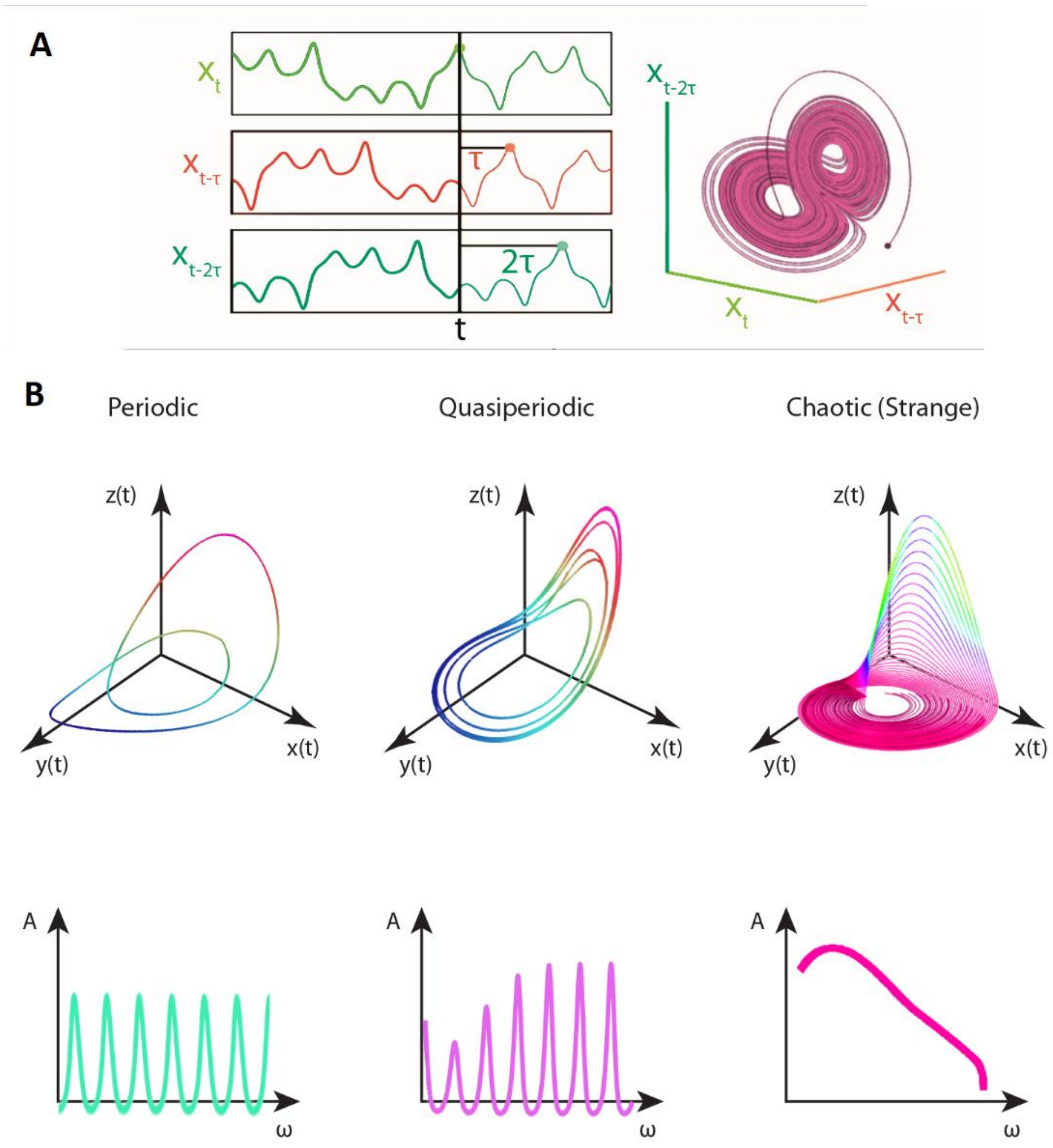

**FIGURE 1. ATTRACTORS AND OSCILLATIONS.** *A) Time-delay Coordinate Embedding. A schematic of attractor reconstruction from a time-series signal of some variable X(t) is shown by time-delay embedding (i.e., Convergent Cross Mapping).* $\tau$ *represent the time-delay. However, for complex large-scale datasets, machine learning algorithms such as reservoir computing (RC) and deep learning architectures are suggested (Image was adapted from Sugihara et al., 2012). B) Three different types of attractors which can self-organize in the signaling/expression state-space of cancer processes are shown: a limit cycle (periodic oscillation), quasi-periodic attractor, and a strange attractor (chaotic). The simplest of attractors, a fixed-point, is not shown herein. Their corresponding frequency spectra are shown below,*

*with the oscillator's angular frequency as the independent variable and the amplitude of the oscillations as the dependent variable. The oscillation of a limit cycle attractor has a defined amplitude (A) and peak in the frequency spectrum at a frequency* $(\omega)$. *A broad frequency spectrum is observed for the strange attractor, which exhibits a fractal-dimension in state-space. However, the frequency/power spectrum can be more complex depending on the system. For instance, complex attractors, such as those observed fluid turbulence, exhibit a broad frequency spectrum with an anomalous power-law scaling (i.e., multifractality) due to intermittency.*

In the chaotic and complex regimes of dynamical systems, multiple attractors may self-organize for a wide range of initial conditions. Some attractors may entangle with those nearby to form complex webs of attractors. Attractors can be of four major types: fixed-points, limit cycles (periodic oscillators), quasi-periodic (tori), and chaotic (strange attractors) (Figure 1A). The first two are equilibria attractors and may correspond to stable oscillations or stable cell phenotypes. The chaotic (strange) attractor is the most complex of attractors and can only be reconstructed in a three-dimensional (or greater) system. They are attractors with a *fractal dimension* (Grebogi et al., 1987). Hyperchaos and hyperchaotic attractors may also emerge in higher-dimensional systems. Hyperchaos, as first coined by Rössler, describes chaotic systems in which more than one positive Lyapunov exponents are observed. If a system exhibits hyperchaos, its three-dimensional projection of the strange attractor may appear noisy and random-like (Letellier and Rössler, 2007). However, in four dimensions or higher, a well-defined strange attractor may emerge. Although the transition of chaos to hyperchaos will not be discussed herein, readers are reminded that certain complex systems may exhibit strange attractors which can only be defined in four-dimensions or higher but appear like noisy, random systems in three or lower dimensions (Letellier and Rössler, 2007).

We need three dimensions (or higher) to analyze chaotic attractors, and hence, observing a single protein oscillation in time as a one- dimensional system is insufficient to detect chaotic behavior (Strogatz, 2015). As such, the most effective classical method for chaotic behavior detection is to embed the time-trace signal onto state-space by Takens' theorem and quantify its Lyapunov exponents and fractal dimension. *Time-delay embedding* allows us to reconstruct a higher dimensional space of the protein flows or gene expression. However, there may be still smearing by noise. Denoising algorithms can be used as a filtering and pre-processing step prior to the attractor embedding. Denoising algorithms can be of many sorts from basic normalization techniques to wavelet analysis and imputation techniques, used for noise reduction in the dataset. From the time-embedded signaling data, one can identify two points that are very close in the phase space and subsequently measure the initial, exponential separation away from each other (Strogatz, 2015). This procedure determines the spectrum of Lyapunov exponents, where the presence of one or more positive Lyapunov exponents imply chaotic behavior (Strogatz, 2015). The presence of a chaotic attractor is further confirmed by assessing the embedded attractor's fractal dimension in state-space (Strogatz, 2015). However, when large complex datasets in the order of thousands of genes or proteins within thousands of cells are considered, as is the case for patient-derived tumor/liquid biopsies and their single-cell analyses, the application of these embedding techniques and traditional chaotic behavior measures may not be sufficient due to dimensionality constraints. Therefore, as will be discussed, machine learning algorithms and algorithmic information dynamics are suggested as robust tools for mapping complex dynamics and inferring chaotic behavior in larger multidimensional datasets pertinent to systems oncology.

Chaotic behavior implies long-term unpredictability, irregularity, and complexity, making the disease difficult to treat. A healthy cell phenotype may correspond to a stable attractor in state space such as a fixed-point or a limit cycle (oscillations). However, a cell if found to be a chaotic attractor, is an unstable state with irregular and aperiodic signaling dynamics. Chaotic dynamics in certain biological oscillations may be robust biomarkers or patterns for diseases. An emerging paradigm in the study of complex diseases, such as cancers, is that chaotic dynamics can emerge in biological oscillators such as gene and protein networks. According to mathematical and computational models, the emergence of chaotic attractors in complex cancer processes have been suggested as indicators of therapy resistance, cancer relapse, emergence of aggressive phenotypes, increased phenotypic plasticity, and metastatic invasion (Itik and Banks, 2010; Khajanchi et al., 2018; Heltberg et al., 2019). However, most of these studies were limited to cell population dynamics. Further, Huang et al. (2009) were amidst the first to suggest using transcriptomic analyses that cancer cell fates are aberrant, embryonic-stem cell like attractors of the Waddington developmental landscape (Zenil et al., 2019). Further, they suggested that cancer stem cells occupy higher energy states of the landscape, thus representing more complex attractors with higher differentiation potency (Huang et al., 2009).

Chaotic dynamics at the level of protein and gene oscillations may also confer dynamical heterogeneity and adaptive survival in individual cell states to withstand extreme environmental conditions due to their large signaling fluctuations (Heltberg et al., 2019). As such, it is further suggested here *chaotic attractors may be signature hallmarks of cancer stemness.* Measuring chaotic attractors then provides a solution to forecast the complex adaptive behaviors and dynamics of the disease system. The chaotic attractor provides a control system framework to reprogram the disease state dynamics in signalling state-space. If chaotic behavior emerges or is at the origin in cancer signaling dynamics, two fundamental questions arise: (1) How do we detect chaotic behavior of or in cancer cells? More specifically, how do we detect strange attractors in the (multiomic) signaling state-space of cancer networks? and (2) How do we distinguish chaotic oscillations from stochastic oscillations (randomness) or noise in the signaling state-space? To address these questions, a brief survey of tools and methods to detect strange attractors in cancer signaling state-space are outlined in this paper. While chaotic oscillations may occur in any cancer-related process, we will primarily focus our attention to the complex networks steering cancer cell fate dynamics and differentiation processes.

**Summary:** Complex systems theory was elucidated as the study of complex processes, patterns, and behaviors in complex systems. An overview of complexity science and the general features/properties of complex systems pertinent to the study of cancer dynamics were reviewed in this section. The emergence of causal patterns known as attractors to which the dynamics of molecular processes or cellular dynamics are bound to in state-space was introduced. Reconstructing the attractors underlying cancer differentiation dynamics and complex cancer processes in gene expression or signaling state-space is a recurrent theme of the paper's causal inference approaches.

## 3. OSCILLATIONS AND CELL PATTERNING SYSTEMS

Given the reconstruction of the complex network patterns/dynamics steering cancer cell fate decision-making, we must understand how their regulatory feedback loops behave in time (i.e., oscillations). As mentioned, cells, genes, and proteins, can be treated as physiological oscillators. Thus, a brief intuition for oscillations is required to understand the use of chaotic-behavior detection tools, in the context of forecasting cancer cell signaling/dynamics. Cancers are essentially characterized by their abnormal, uncontrolled cell division due to chromosomal/genomic instabilities. The tumor suppressor transcription factor p53 is highly conserved at the protein level and plays a key role in DNA damage response. It is a master regulator of the cell cycle and hence, cell proliferation. In cancer cells, the TP53 gene is mutated (loss of function) in about 50% of all cancers and serves as the critical bifurcation point for tumorigenesis on the cell developmental/differentiation landscape (Bar-Or et al., 2000). The core p53-MDM2 negative feedback loop shows that the synthesis-degradation kinetics of p53 and MDM2 governs their oscillations within cells. The oscillations of this regulatory feedback mechanism are essential for the signaling dynamics of all other intertwined proteins and genes regulating cell homeostasis and cell cycle control. Within this circuit, p53 transcriptionally activates *mdm2*. Mdm2, in turn, negatively regulates p53 by both inhibiting its activity as a transcription factor and by enhancing its degradation rate. Models of negative feedback loops, such as between p53 and Mdm2, suggest that they can generate an *oscillatory behavior* due to a time delay between the two proteins activity (conformation states). However, with DNA damage (as in the case of tumors harboring∼ hundreds of mutations), excessive continuous p53 oscillations are observed (Bar-Or et al., 2000).

Fluorescently tagged fusion proteins can be used for the time-lapse imaging of these proteins and quantify their oscillatory dynamics within cells. For different parameters of the feedback loop, the dynamics can show either a monotonic response, damped oscillations, or undamped oscillations. The stronger the coupling interactions between the proteins, the more oscillatory the dynamics tend to be. The kinetic parameters act as damping or signal-amplifying coefficients. For instance, high basal degradation rates of the proteins act as damping coefficients of the oscillations. To illustrate an experimental approach to measuring cellular protein oscillations, Geva-Zatorsky et al. (2006) experimentally investigated fluorescently labelled p53 and Mdm2 dynamics in living breast cancer (MCF7) cells with perturbation analysis by gamma irradiation. In a large fraction of cells, they found undamped oscillations of p53-CFP and Mdm2-YFP, which lasted for at least ∼3 days post- gamma irradiation. Aside from the noise fluctuations in cell-cell variability, two characteristic properties were found in the protein oscillations: (1) the oscillation amplitudes fluctuated widely, yet the oscillation frequency was much less variable, and (2) while some cells regularly oscillated, other cells showed a dynamic fluctuation of protein levels (i.e., signaling heterogeneity) indicating the presence of complex, irregular oscillations. Essentially, the negative feedback loop amplified slowly varying noise in the protein production rates at frequencies near the resonance frequency of the feedback loop. However, none of these studies performed chaotic behavior detection such as time-delay embedding of the signal followed by calculation of Lyapunov exponents nor computing the fractal dimension of observed attractors. The description of the fluctuations and increases in cellular oscillations were qualitative in most part.

Similarly, many embryonic developmental factors (i.e., morphogens) involved in cellular patterning systems are known to exhibit oscillatory dynamics. The oscillatory states of these signals show typical time periods of a few hours (1-4 hrs) and are referred to as ultradian oscillations (Jensen et al., 2010). A key example of ultradian oscillations is the coupling of Wnt and Notch signaling (Sonnen et al., 2018). One of the theoretical frameworks for understanding these protein/gene oscillations in developmental pattern formation has been laid by the clock-and-gradient, or clock-and-wavefront model, originally proposed by Cooke and Zeeman (1976). The dynamic signal encoding based on relative timing of oscillatory protein signals are essential for the development of the embryo. It has been shown that a phase-shift in morphogen oscillations fine-tune segmentation in the developing embryo (Sonnen et al., 2018). These morphogens are aberrantly expressed in cancer stem cells and are key signaling factors of the cancer stem cell niche (Plaks et al., 2015; Batlle and Clevers, 2017). However, the experimental study of morphogen oscillations via time-lapse imaging within pathological cell states (such as cancer stem cells) remain nearly absent. Their oscillatory dynamics have not been investigated experimentally in time-series cancer stem cell differentiation or during cancer cell fate decisions such as proliferation and differentiation dynamics. Moreover, the detection of quantitative cellular behavioral techniques as advocated herein are virtually unknown to most cancer researchers investigating protein-mediated pattern formation. Therefore, the lack of experimental time-series cancer datasets (due to technological limitations) and a lack of dynamical systems theory applications in cancer research go hand in hand and limit our understanding of how complex dynamics may be orchestrating tumor patterning systems.

As explained, when the amplitude of an oscillation by coupling to some external signal increases beyond a critical threshold, chaotic dynamics can emerge as indicated by aperiodicity/irregularity in the oscillations or period-doubling bifurcations. To illustrate, the Mackey-Glass equations have shown that variations of chemical concentrations, such as the production rates of proteins within cells or the cell-density variation, may exhibit a time-delay (Mackey and Glass, 1977). An increase in production rate k caused by a time-delay $k(t-\tau)$ may result in pathological diseases. Time-delays are control parameters which above a certain value may result in chaotic cellular oscillations (Mackey and Glass, 1977). According to Jensen et al. (2012) a negative feedback loop is a sufficient requirement for chaotic behavior to emerge within cell signaling. A negative-feedback loop ensures the presence of a time-delay in cellular oscillations, and thereby may act as precursor for the onset of chaotic dynamics (Jensen et al., 2012).

There are also theoretical works by Kaneko and Furusawa exploring the cancer stem cell hypothesis through the lens of cell adhesion and oscillatory nonlinear dynamics which warrant further investigation (Furusawa and Kaneko, 2001; 2009; 2012). In their chaos hypothesis, they propose that the robustness and differentiation of stem cells towards their multipotent complex cellular states can be predicted by chaotic intra-cellular chemical dynamics (Furusawa and Kaneko, 2001). Intracellular chaotic oscillations were suggested as markers of pluripotency (stemness) (Furusawa and Kaneko, 2012). Dysregulated focal adhesion dynamics to the extracellular matrices are hallmarks of cancer metastasis and EMT state-transitions/plasticity dynamics. Most cancer-related deaths are caused by metastatic invasion, and hence elucidating the nonlinear dynamics underlying such plasticity transitions/differentiation dynamics may help identify causal markers in controlling and regulating their behavioral patterns.

While mathematical models, such as differential equations with time-delays, may in principle capture strange attractors within cellular protein flows and gene signaling, experimental confirmation of intracellular chaos remains a fundamental roadblock in complex systems research. Therefore, the quantification of protein oscillations, using the techniques described above must be performed in cancer cells and CSCs in time-series and subjected to the various behavioral detection methods enlisted herein. For instance, we can identify the negative feedback loops regulating the Suvà glioma stemness network (*POU3F2*, *SOX2*, *SALL2*, and OLIG2) and experimentally quantify their oscillations using time-lapse fluorescent-imaging within GBM-derived cancer stem cells to understand their cell fate dynamics and reprogrammability (Suvà et al., 2014). Without such experimental datasets, the plausibility of chaotic dynamics as a hallmark of cancer signaling/progression cannot be verified and computational/systems oncology will remain bound to computational and mathematical models. Having laid the basic intuition for cancer stemness and the complex feedback loops regulating their oscillatory dynamics, the following methods are discussed as approaches to detecting chaotic behavior and complex dynamics in cancer networks given their time-series signals are made available.

**Summary:** An intuition for intracellular oscillations and multiscale pattern formation in cancer systems was provided in this section. The concept of feedback loops and time-delays in intracellular oscillations was illustrated in relation to the emergence of complex attractors in their corresponding state-space. The oscillatory dynamics of proteins, genes, cells, and other systems were discussed as information dynamics of underlying graph-theoretic networks.

## 4. TAKENS' THEOREM

Let us suppose the flows of protein densities and gene expression dynamics of cancer (stem) cells are available in time-series, and a normalized gene expression (or protein oscillation) matrix, array of genes (proteins) by cells with their count, is produced for each time-point. Takens' time-delay coordinate embedding is the state-of-the-art approach for the attractor reconstruction underlying these complex signals. In 1981, Takens demonstrated in his embedding theorem that the topological dynamics of a complex multidimensional system can be derived from the time-series of a single observable variable (Takens, 1981). The embedding theorem was first demonstrated in the study of fluid turbulence. The term *strange attractors* were coined by Ruelle and Takens to describe the multifractal patterns observed in the bifurcations of turbulent fluid flows (Ruelle and Takens, 1971; Gleick, 2008). They defined a strange attractor as the local product of a Cantor set and a piece of a two-dimensional manifold (Ruelle and Takens, 1971). Takens' theorem can be used to embed the three-dimensional flows of not only turbulent flows but in principle, any chaotic signal in state-space.

According to Takens' theorem, you can roughly reconstruct the state-space of a dynamical system by delay-embedding only one of its time-series projections, given the assumption that the variable X contains redundant information about Y and Z variables (i.e., they are causally- related) (See rEDM link in Appendix). From the perspective of Shannon's information theory, the optimal time-delay $\tau$ to reconstruct the state-space attractor would correspond to the minimum Mutual Information (MI) of the system (Fraser and Swinney, 1986). The complex structure obtained by the embedding is the *attractor*. As discussed, in chaotic systems, an attractor with a (multi) fractal dimension is observed, known as the strange attractor to which the trajectories of the chaotic system are bound to. Thus, once an attractor is obtained from the time-series embedding, the fractal dimension and Lyapunov exponents can be computed to assess the stability of the dynamical system and verify if the identified attractor is a *strange attractor*. This time-delay embedding procedure scales with the time-series. The longer the time-series, the larger the network and the more complex the attractor obtained. Perturbation analysis can assess the stability and robustness of the complex attractor and provides a powerful toolkit for complex networks analyses in Algorithmic Information Dynamics (AID), as will be discussed later.

An rEDM package for time-delay embedding on experimental time-series is provided in the Appendix. rEDM is an R-package for Empirical Dynamic Modelling and Convergent Cross Mapping (CCM) as devised by Sugihara et al. (2012). The causal relationships in complex disease signaling networks can be identified using CCM (Krieger et al., 2020). The rEDM package uses a nearest neighbor forecasting method with a Simplex Projection, to produce forecasts of the time-series as the correlation between observed and predicted values are computed (Sugihara et al., 2012). CCM is an embedding technique which combines Takens' theorem and Whitney's embedding theorem. CCM measures the extent to which states of variable Y(t) and Z(t) can reliably estimate states of variable X(t), as explained above. This happens only if X(t) is causally influencing Y(t) and Z(t). There is a simpler R-package called 'multispatialCCM' (multi-spatial Convergent Cross Mapping), an adaptation of CCM, for chaotic time-series attractor reconstruction available, as well. CCM (Takens' theorem) should be amidst the first set of causality-inference algorithms used to embed the time-series cancer signals in state-space and reconstruct their underlying attractors. However, there may be dimensionality limits for such approaches when dealing with multi-dimensional complex systems like the cancer transcriptome. These embedding methods may be useful for trajectory inference in cell population dynamics or for brute-force approaches in cellular signaling. For instance, we can embed the time-traces of proteins or signals with predicted chaotic dynamics from literature analysis. Otherwise, the task may be too complex for these methods, and non-traditional detection tools must be employed (i.e., machine intelligence).

**Summary:** A traditional algorithm/method for attractor reconstruction from time-resolved signals/data was discussed. Takens' theorem and convergent cross mapping were presented as the most basic of time-delay embedding techniques for identifying causal structures and patterns in cell fate dynamics or intracellular cancer oscillations. However, such methods may be restricted to dimensionality limits and may not be applicable for complex time-series datasets, such as those pertaining to many-body, multiscale gene/protein signaling networks. Regardless, these tools can still be exploited in the data science of simplified networks such as simple transcriptional/gene circuits or few protein oscillation systems.

## 5. LYAPUNOV EXPONENTS

Once the chaotic signal has been time-embedded and its attractor(s) has been reconstructed in state-space, quantitative-behavioural detection algorithms can be used to determine if the identified attractor is indeed chaotic (strange). As such, Lyapunov exponents and fractal dimension estimates remain the most robust classical measures of chaotic behavior detection. Positive Lyapunov exponent(s) are robust signatures of chaos quantifying *sensitive dependence on initial conditions*. Prior to measuring the Lyapunov exponents or fractal dimension on the identified attractor, de-noising and filtering algorithms can be used as a pre-processing step to reduce the noise and data dispersion. The intuition behind this is to consider nearby points in the phase space generated by time-embedding and then perturbing each point proportional to a weighted average of the nearby points. Using this, one can recover the fine structure of the attractor in higher dimensions, especially if this is combined with signal smoothening methods (e.g., imputation algorithms).

To illustrate Lyapunov exponents, consider two points of a cancer signal's trajectory (e.g., a single protein flow or gene expression time-trace) separated by a very small distance in time, x(t), in phase space, then the separation (bifurcation) of its trajectory from its initial position is given by:

$$\delta x(t) \approx x_0 e^{\lambda_L \mathrm{t}}$$

for a small time, t, where the $\lambda_L$ is the Lyapunov exponent. In multi-dimensional dynamical systems, there may be a spectrum of Lyapunov exponents to consider. The Lyapunov exponent measures how far two initially close by points on a dynamical system's trajectory separate (bifurcate) in time. If the Maximal Lyapunov exponent $\lambda_L > 0$ (positive), there may be a chaotic attractor (Strogatz, 2015). That is, in a chaotic system, the trajectories exponentially diverge apart from each other. In hyperchaotic systems, at least two positive Lyapunov exponents are observed. The inverse of the positive Lyapunov exponent $\frac{1}{\lambda_L{}^+} \sim \tau_L$ denotes the Lyapunov time $\tau_L$, the finite predictability horizon of a chaotic system. However, unlike a random (stochastic) system, the exponentially diverging trajectories will map onto a finite fractal structure in phase space: the *strange attractor*. As mentioned, the fractal topology of chaotic attractors provides the explanation for this counter-intuitive pattern. The fractal topology allows the *stretching and folding* of phase-space analogous to making taffy candies (Gleick, 2008). The NP-hard question then is how to find these taffies in the state-space of a given complex system/network's state-space?

Lyapunov exponents have been used in mathematical and numerical simulation models of cancer population dynamics to verify the presence of chaotic attractors. For example, Itik and Banks (2010) demonstrated using a set of ordinary differential equations (ODE) to model the tumor-immune-host cell density dynamics the emergence of chaotic attractors at certain critical parameters such as the growth rate. The existence of chaotic behavior was confirmed by calculating the Lyapunov exponents and the fractal dimension of the observed attractor, which was found to be near that of the Lorenz attractor's fractal dimension. Thus, the pairing of computational models and simulations with empirical dynamics is fundamental to complex systems research. Chaotic dynamics were confirmed in the model by a positive maximum Lyapunov exponent. As the $\alpha$ parameter increased, transition to chaotic behavior was observed in the bifurcation plot by period-doubling cascades, signatures of chaotic dynamics (Itik and Banks, 2010).

More recent mathematical modelling of tumor-immune cell predator-prey dynamics have been performed with time-delay as a bifurcation parameter (Khajanchi et al., 2018). Chaotic attractors emerged in the system's phase-space and were suggested as indicators of aggressive metastatic transition in cancer cells (Itik and Banks, 2010; Khajanchi et al., 2018). However, as stated, the lack of time-series experimental datasets remains a roadblock in experimentally confirming the presence of these chaotic attractors in cancer-immune-host dynamics. As such, a python and MATLAB code for Lyapunov exponents calculation from time-series is provided in the appendix to encourage its applications in medical systems. The 'nolds' package in python, a small numpy based library, also provides various measures of nonlinear dynamics such as the Lyapunov exponent and Hurst index (see Appendix).

**Summary:** Positive Lyapunov spectra/exponents was discussed as a robust signature of chaotic dynamics. Methods to quantify or predict the Lyapunov spectra of complex multiscale cancer signals/datasets, including gene expression, protein oscillations/transcriptional circuits, and imaging analyses were presented herein.

## 6. FRACTAL DIMENSION AND MULTIFRACTAL ANALYSIS

Fractals are ubiquitous in nature. They are universal patterns exhibiting self-similarity which iterate themselves across many scales (i.e., power law scaling). We tend to think of trees, snowflakes, clouds, blood vessels, bronchi, neural networks, or the coastlines of geographic landscapes when thinking of fractal structures (Mandelbrot, 1982). They demonstrate that complex geometric patterns can spontaneously emerge from simple recursive rules. Chaotic attractors and many complex systems exhibit (multi)fractal scaling. A such, fractals serve as a robust measure of both, complexity, and chaotic dynamics. However, there is a caveat. The complexity we refer to here is not algorithmic complexity, but rather complex dynamics and irregularity. From the viewpoint of algorithmic complexity, only a very short program is required to generate fractal patterns. Hence, there is some ambiguity between complex fractal dynamics and algorithmic complexity measures, in this domain of research. Fractal geometry explains the paradox of how strange attractors compactify infinite curves into a finite space (area or volume), and as such we may define this recursive irreducibility as complex dynamic structures.

The word fractal, coined by Mandelbrot, is derived from the Latin word *fractus* meaning *fragmented*. Formally, a fractal is defined as a mathematical object with a fractional (non-integer) dimension (Mandelbrot, 1982). Gaston Julia first demonstrated that iterated functions of complex numbers can generate fractal patterns. However, Mandelbrot computationally generated fractals and demonstrated their universality in Nature and chaotic systems (Mandelbrot, 1982). The simplest example of a fractal is the Cantor Set (dust) in which one starts with a straight line and as we keep removing the middle one-third of the line with each iteration, the fractal is generated. Another set of examples are the Serpinski's gasket and Koch's snowflake, both generated by simple recursive rules starting from an equilateral triangle. Fractals can also be continuous. A good example would be Hilbert's space-filling curves which remind us of Escher's tessellations or the honey-comb lattices in beehives. However, the most popular example of a fractal is the Mandelbrot Set, the set of all Julia sets, described by the iterative complex function z defined at n-iterations by the rule: $z_{n+1} = z_n + c$, where the complex number c is its initial condition $z_0$. The Mandelbrot set demonstrates a central property of many complex systems, that complex structures and patterns can be generated from simple, recursive feedback loops.

Fractals are some of Nature's most stable structures demonstrating the optimization of space (compactification) and its spatial resources. A system exhibiting fractal architecture is robust to environmental changes. For examples, bees use a fractal space-filling architecture, composed of hexagonal symmetry to optimize the area: curve length (perimeter) ratio in building their beehive structures. The same principles of hierarchical spatial organization and resource optimization may apply to Nature's intelligent exploitation of fractal geometry, as an adaptive strategy to minimize the amounts of resources used and wastes produced by a complex system (Mandelbrot, 1982). For instance, oil spills exhibit fractal patterns in ocean floors and lakes (Benelli and Garzelli, 1999). The fractal dimension can be computed from their imaging power spectra (the ratio between powers at different scales) to characterize their texture analysis (Benelli and Garzelli, 1999). Their fractal structure may imply that they are difficult to treat as their patterns and information repeatedly span across many scales. Similarly, studies have shown that tumor textures can be characterized as multifractal structures (Baish and Jain, 1998; 2000; Lennon et al., 2015). In the study of tumor structures, fractals have been restricted to describing the self-similarity of abnormal blood vessels (angiogenesis) and tumor contours across many length scales, as a measure of its spatial roughness (Baish and Jain, 2000). However, it can also be applied to time-series analysis as well, as in the case of detecting strange attractors from complex signals. The multifractality of tumors may reveal their aggressiveness, resilience (to environmental perturbations) and hence, be indicators of tumor relapse and therapy resistance (Itik and Banks, 2010). Mesoscopic mathematical models of tumor pattern formation dynamics have also suggested that a fractal dimension analysis could provide a quantitative measure of its growth forecasting and irregularities (Izquierdo-Kulich et al., 2008; Izquierdo-Kulich et al., 2009).

The Fractal Dimension (FD) is a statistical measure of complexity which occupies a fractional dimension, in between two consecutive integers. That is, if a point has a dimension of zero, a line is 1-D, a plane is 2-dimension and volume is 3-D, fractals occupy dimensions in between these integers. FDs can be used to characterize the structural complexity (roughness) of tumors and their irregularity in their signaling dynamics (time-traces). For example, the tumor vasculature was shown to have a higher fractal dimension of $1.89 \pm 0.04$, whereas normal arteries and veins yield dimensions of $1.70 \pm 0.03$ (Baish and Jain, 1998; 2000). The higher FD indicates an increased roughness and complexity of the vasculature. The fractal dimension of an image, such as medical imaging of tumor structures, may be estimated by various techniques: (*a*) box-counting/cube-counting (for volumetric systems); (*b*) correlation; (*c*) sandbox; (*d*) Fourier spectrum, etc. When applied to images of blood vessels, these methods yield scaling relationships that are statistical best fits to a power-law relationship within a finite range of scales. Although, these fractal dimension estimation algorithms differ, they obey to the same calculation basis summarized by the three steps: (1) Measure the quantities of the object using various step sizes, (2) Plot log (measured quantities) versus log (step sizes) and fit a least-squares regression line through the data points, and (3) Estimate FD as the slope of the regression line (Grebogi et al., 1987; Strogatz, 2015).

The most widely used FD computing algorithm is the Box count algorithm. In the Box counting method the signals are represented on a finite scale grid and the grid effects interplayed with the computing fractal dimension. The box-counting method asks: How many boxes are needed to cover the fractal? A fractal can be described by a power law scaling given by $N \propto r^{-D}$ where N is the number of boxes needed to cover the object/pattern, r is the length of the box, and D is the fractal dimension. Then, $D = \frac{logN}{\log\frac{1}{r}}$, where 1/r is the inverse of the box size r (Grebogi et al., 1987). Therefore, the slope of the log-log plot of N and r is the Fractal dimension. However, what if the system requires more than one FD to characterize its statistical patterns? Complex dynamical systems may exhibit multifractality when there are scaling processes in time. The time-series power (frequency) spectra of fluid turbulence exhibit intermittency and fluctuations, which necessitate the use of multifractal analysis (Ruelle, 1995). There may be hidden spikes (sudden transitions) in the intermittent fluctuations of experimental fluid turbulence (Ruelle, 1995). Multifractal analysis provides a powerful tool to characterize these fluctuations in complex dynamical systems.

Multifractal analysis was first introduced by Mandelbrot in the study of turbulence-mediated flow velocity patterns. The Multifractal spectrum can be quantified by the following descriptors: (a) the Hurst exponent, (b) the slope of the distribution produced by the collection of the Hölder regularity index $\alpha$, and (c) the width spread (broadness) of the spectrum, characterizing the variability of the Hölder exponents (Lopes and Betrouni, 2009). The local Holder exponent $\alpha$ is a local measure of roughness, and an exponent of a power law characterizing the multifractality of the system. As the word *spectrum* implies a multifractal is a process exhibiting scaling for a range of different power laws. Various methods exist for computing the multifractal spectrum of Holder exponents (i.e., the slope of the log-log plot of the power law system), which include fractional Brownian motion (fBm) methods, and the most popular are the wavelet-transform based methods. For instance, the Wavelet Transform Modulus Maxima (WTMM) method, uses the continuous wavelet transform to compute the Hölder exponents. It is more efficient than the box-counting algorithm. The local Holder exponent is defined as:

$$d_{loc}(x,y) = \lim_{n\to\infty} \frac{\log\big(Prob(i_1 \dots i_n)\big)}{\log\,(2^{-n})}$$

Where the Prob (…) term is the probability that the point (x,y) of the signal lies in a square with indices ($i_1 \dots i_n$). N addresses the number of squares containing (x,y). As we take the limit defined, we get the multifractal spectra a collection of all points of the fractal having the local Holder exponents alpha (i.e., $d_{loc}$ becomes alpha) (Yale University, Multifractals). The Hurst index, H, describes the roughness of the time-series. It takes a value in between 0 and 1 wherein H= 0.5 denotes a true random process (i.e., Brownian time-series). The smaller the value of H, the higher the roughness, and vice versa.

**Summary:** Fractal dimension and multifractality were discussed as signatures of chaotic dynamics. The notions of recursiveness, modularity, statistical self-similarity, and roughness/irregularity in chaotic dynamics was illustrated. Tools for quantifying the fractal index were discussed as methods to capture chaoticity in the behavioral patterns of collective cancer cell fate decisions/differentiation dynamics. Further, we made a distinction between chaos and regularity instead of chaos and order, since chaotic dynamics itself is a complex causal order.

## 7. CRITICALITY

A feature of many robust complex systems is *criticality* (i.e., edge of chaos). The transition from criticality to chaotic dynamics may be most useful for our discussion of cancer cell fate dynamics. Critical systems are a class of nonequilibrium systems exhibiting scale-invariant spatial-organization and scale-free dynamics (Christensen and Moloney, 2005). In nonequilibrium systems, the critical points indicate regions where the attractors governing its phase-space dynamics are located (Fang et al., 2019). It has long been suggested by Kauffman et al. that gene regulatory networks operate in the critical phase between regularity and chaos. Critical dynamics of the network were suggested to permit the coexistence of robustness and adaptability/resilience in cellular systems, thereby allowing both the stability of cell fates and their epigenetic switching between multiple phenotypes (network states) in response to environmental fluctuations/perturbations and/or developmental cues (Kauffman, 1995). The emergence of phenotypic plasticity may be deep-rooted in critical dynamics in complex networks and chromatin states configuration. However, the lack of time-series datasets limits the observations of critical network dynamics and critical cell-state transitions to computational models such as Boolean networks (Torres-Sosa et al., 2012). Further, the lack of time-series data points has fundamentally limited the investigation of critical dynamics in cellular processes to simulations. Therefore, simulations-driven artificial intelligence (AI) is a powerful platform in complex systems research granting us insights into otherwise intractable problems.

In dynamics, a critical point or tipping point is a state that lies at the boundary between two phases or regions that have differing behaviors and rules. According to Renormalization Group theory, phase transitions are characterized by a divergence (i.e., tends to infinity) in the coherence length (the characteristic length scale) of a system near some critical point (Bossomaier and Green, 2000). The corresponding quantity of interest in a dynamical system is the correlation time, the length of time for which perturbations are propagated into the future. Critical systems exhibit *power law* behaviors as signatures of their long-range interactions in the system. Criticality may be an indicator a complex system's potential to *transition to chaotic behavior* (Bossomaier and Green, 2000; Sethna, 2006). As such, the detection of power law statistics can be indicators of complex dynamics, and studying criticality reduces the search space of potential chaotic oscillators within a complex network.

Critical dynamics have been modelled in the cell fate transitions (differentiation dynamics) from healthy to cancer cell states (Rockne et al., 2020). Rockne et al. (2020) demonstrated the state-transition dynamics from healthy peripheral mononuclear blood cells (PBMC) to acute myeloid leukemia (AML) in mice can be described by a double-well potential with two critical points. They demonstrated the critical points in the transcriptomic state-space can predict the cell fate trajectories during disease progression. The 2D transcriptomic state-space was obtained from dimensionality reduction analysis (PCA- principal component analysis) on the time-series bulk RNA-seq data (Rockne et al., 2020). The transcriptome was modelled as a particle undergoing Brownian motion using the Langevin equation in a double-well quasi-potential U(x) with two stable states, the critical points, representing the healthy and AML states, respectively. To calculate the mean expected stochastic behavior of the cells, they considered every point in the transcriptome state-space as a particle characterizing a cell. The evolution of the probability density function of all such particles (cells) was obtained by the Fokker–Planck equation (FPE) (Rockne et al., 2020). Although, the model beautifully illustrates critical dynamics in cancer cell fate transitions, the assumption that the healthy state and AML state critical points are *stable phenotypes* may be the issue the issue with the model. The cell states were assumed to be *stable fixed-point attractors* separated by an unstable transition-state higher in potential energy. As discussed herein, cell phenotypes may correspond to various types of attractor patterns (i.e., fixed-points, limit cycles, tori, chaotic) on the multidimensional signaling state-space. However, one-dimensional dynamical systems, as modelled in this study and most studies in systems oncology, are limited to fixed-point attractors only.

Cell fate bifurcations may exhibit critical dynamics. A key mechanism for cell fate transitions in cancer systems is the EMT program (Epithelial-Mesenchymal Transition) and remains as one of the primary examples of critical cell fate dynamics observed in the computational models by Nieto-Villar et al. and Jolly et al. EMT programs govern many cancer-related behaviors such as the transition from one cancer phenotype to another, stem cell plasticity, cancer metastasis and chemoresistance. EMT switches are essential in the critical dynamics of CSCs and non-CSC phenotypic switching (Plaks et al., 2015; Batlle and Clevers, 2017).

EMT transitions have been modelled as first-order phase -transitions, a signature of critical systems, in computational tumor dynamics models by Guerra et al. (2018). Cancer evolves along three basic steps: avascular, vascular, and metastatic, all emerging downstream of biological phase transitions. Guerra et al. (2018) demonstrated a network model of EMT dynamics consisting of four interacting cell types, wherein N represented the population of normal cells exposed to pro-carcinogenic stimulus, H the healthy cell population (mainly epithelial phenotype), and M is the population of mesenchymal cells (Guerra et al., 2018). The immune population I was used as the control parameter (can fluctuate). The network consisted of various cellular processes including mitosis and apoptosis of the proliferating tumor cells. Mathematical models of chemical kinetics were used to reduce the network to a system of ODEs representing the EMT dynamics. Lempel-Ziv compression algorithm, Lyapunov exponents, and the Lyapunov fractal dimension were assessed on the computational model dynamics. At some threshold of the control parameter I, EMT was observed in the dynamics as characterized by a supercritical Andronov-Hopf bifurcation and emerge of a limit cycle (Guerra et al., 2018). As the control parameter I further decreased below critical thresholds, complex Shilnikov-bifurcations were observed, and the population dynamics eventually gave birth to chaotic dynamics. The computational model exhibited that under decreased immune dynamics (i.e., lower I value), the tumor cells exhibited apparently random behavior (i.e., chaotic dynamics), thus promoting mesenchymal phenotypes as indicated by the EMT phase-transition (Guerra et al., 2018).

Further in evidence to critical dynamics in EMT systems, one of the many complex signals mediating EMT transition is the microRNA-200/ZEB mutual inhibitor feedback loop, driven by the transcription factor SNAIL (Sarkar et al., 2019). A simulation-based study found that mRNA levels of ZEB can indicate the critical tipping point for EMT phase transitions in cancer cells (Sarkar et al., 2019). An increased variance, autocorrelation, and conditional heteroskedasticity were shown to dynamically vary during the phenotypic transitions, with an increased attractor basin stability observed for the hybrid EMT state, indicating it may be the fittest phenotype for metastatic progression (Sarkar et al., 2019). The mathematical models by Sarkar et al. showed a cusp-like catastrophe in the EMT plasticity bifurcation diagram, suggestive of a critical phase-transition.

Sarkar et al. (2019) considered a mathematical model of microRNA-based chimeric circuit capturing the binding/unbinding catalytic kinetics of associated protein complexes and transcriptional machineries. Given m is the abundance (number) of mRNA, let n be the abundance of microRNA, and B be the TF protein of interest, then, we have the first-order kinetic equations:

$$\frac{dn}{dt} = g_n - mY_n - k_n n$$

$$\frac{dm}{dt} = g_m - mY_m - k_m m$$
$$\frac{dB}{dt} = g_B mL - k_B B$$

Where g corresponds to the synthesis rates of the respective molecules in subscript, and in particular, $g_B$ is the translation rate of protein B for each m in the absence of n. The k parameters denote the degradation rates of the molecules, Y and L are the n-dependent functions denoting the various effects of miR-mediated repression (Sarkar et al., 2019). These differential equations were computationally simulated using Monte Carlo simulation in which each reaction event is considered as a Markov process (Sarkar et al., 2019). The time and species numbers were updated stochastically by choosing a random reaction event. The miR-200 based chimeric tristable miR-200/ZEB circuit was simulated by casting 10 reaction events as a function of the number of SNAIL molecules. The corresponding Master Equation was simulated with the Gillespie algorithm to obtain the stochastic trajectories from which the critical transitions was identified with a cusp-like phase-transition in the bifurcation plot (Sarkar et al., 2019). Hari et al. (2021) have extended the understanding of these EMT switches by use of used network approaches to investigate cellular decision-making in EMT phenotypic transitions and how they regulate emergent behaviors such as phenotypic plasticity dynamics (i.e., the ability to reversibly switch/transition in between heterogeneous phenotypes) in tumor ecosystems. The study shows that network topology influences phenotype commitments and canalization signatures of the tumor differentiation/developmental landscape. Many other mathematical studies demonstrating phase-transitions and complex dynamics in metabolic tumor growth models with glycolytic oscillations have been established by the Nieto-Villar group, which shall not be discussed herein (Izquierdo-Kulich et al., 2013; Betancourt-Mar et al., 2017; Martin et al., 2017).

The above-listed studies show that computational simulations/modelling paired with complex networks analysis pave fruitful insights into many other complex cancer processes. There are various other tools borrowed from nonequilibrium statistical physics one can use to investigate criticality and phase-transitions in complex systems. The breadth of this topic deserves a separate paper of its own and cannot be confined to this brief survey. However, the gist of these approaches are summarized below by two key techniques that may be useful for cancer research: percolation clustering, the Ising spin glass model, and Cellular Automata (CA).

Cancer networks can be visualized as Boolean networks, in which the elements of the networks, such as gene expression, can either be on or off, described by 1 and 0, Boolean states. This allows them to be ideal models for the Ising model adaptation, where the Boolean states correspond to gene-type spins (spin up and spin down). Further, many complex cancer processes including phosphoproteomics (on/off switches of protein conformations) and epigenetic-chromatin states (e.g., acetylation/methylation dynamics of histones) can be defined as binary states. As such, Ising models are simple yet powerful tools to study cell fate transitions from such complex gene and protein network dynamics (Mezard and Montanari, 2009; Krishnan et al., 2020). However, as mentioned, even the optimization of a two-dimensional Ising model is NP-hard. Therefore, mean-field theory approximation or iterative random searching algorithms such as Monte-Carlo methods (e.g., the Metropolis algorithm) are used to find approximate solutions. For example, Lang et al. (2014) using single-cell data demonstrated the rugged energy landscapes of Ising models can be used to visualize/reconstruct the distinct cell states from transcriptomic data as attractors of the Waddington epigenetic landscape (Lang et al., 2014). The flipping of spins caused the step by a step-change in the phenotype of the cell, mapping cell fate transitions from one attractor to another. Ising models can predict the co-existence of structurally or functionally organized clusters in complex networks, as well. Ising models also serve as the theoretical framework of Artificial Neural Networks such as the Hopfield neural networks (Hopfield, 1982). Hopfield networks are emerging as machine learning approaches for causal inference in complex multiscale cellular dynamics including classifying or predicting gene expression patterns and forecasting their epigenetic landscapes (Guo and Zheng, 2017).

One of the issues in the understanding of critical dynamics, such as EMT processes in cancer cells, is the lack of a mathematical theory/mechanisms to explain their transition to chaos. The works by Kauffman et al. in this regard have only been qualitative for the most part (Kauffman, 1995). On a tangential note, in literature, one usually distinguishes chaos from "order." This phrasing is extremely common in complex systems research, but it is technically ambiguous. In the 1960s, Prigogine, a pioneer of complexity, demonstrated that disordered, far-from equilibrium chemical systems can spontaneously give birth to orderly stable states (i.e., dissipative structures) (Prigogine and Lefever, 1968). Prigogine defines the self-organization of these dissipative structures as *order out of chaos*. Therefore, *order* may be an emergent behavior of chaotic systems. As such, it is technically more accurate to distinguish chaos from "regularity" or periodicity, rather than "order."

Cellular Automata (CA) are discrete dynamical systems, consisting of a grid/lattice of adjacent cells updated by simple local rules. As mentioned, the Bak sandpile model was a CA which showed self-organized critical dynamics and emergent patterns of behaviors. As such, CA are versatile tools for modelling complex and critical systems (Bak et al., 1987; Fronczak et al., 2006). Further, powerful complex systems frameworks such as tools from Algorithmic Information Dynamics can be coupled with CA to study complex networks dynamics and biological pattern formation (Zenil, 2020).

CA are spatiotemporally discrete patterning systems represented by lattices of local interactions. The transition rules are local and only depend on the site neighborhood interactions on the lattice. Traditionally, at every time step, every lattice suite updates its state simultaneously. However, there are variants of CA such as asynchronous CA and/or inhomogeneous CA where such rules do not apply, as seen in tumor growth models (Moreira and Deutsch, 2002). There is a vast amount of literature on the use of CA to model tumor growth dynamics. Some examples include avascular tumor growth models with the CA system modelling reaction-diffusion equations, partial differential equations (PDE) modelling the tumor-immune-host cell dynamics in varying nutrient conditions (Mallet and de Pillis, 2006). Similar works were seen with inclusion of the effects of chemotherapeutic drugs on the tumor growth in reaction-diffusion PDE models by Ferreira et al. (2003) and in the ODE models of de Pillis and Radunskaya (2001).

In a typical avascular tumor growth model, we have a regular lattice wherein the discrete states correspond to biological (Cancer) cells. If a cell dies, the lattice is unoccupied (Moreira and Deutsch, 2002). A cell can survive, divide, or die, and local rules can be updated asynchronously. We can also simulate the ecological dynamics or competitive interactions between multiple cell types, as seen in heterogeneous tumor microenvironments (Moreira and Deutsch, 2002). For instance, the hybrid PDE-CA simulations by Mallet and Pillis (2006) closely match the population dynamics observed in experimental tumor-immune dynamics. The models can also simulate complex tumor processes such as immune-cell filtration and tumor immune escape (Mallet and de Pillis, 2006). The nutrient species' reaction-diffusion dynamics govern the tumor growth model in these CA systems. Let us consider a simple model system with two nutrients, then the reaction-diffusion system is given by:

$$\frac{\partial N}{\partial t} = D_N \nabla^2 N - k_1 HN - k_2 TN - k_3 IN$$

$$\frac{\partial M}{\partial t} = D_M \nabla^2 N - k_4 HM - k_5 TM - k_6 IM$$

Where, M and N represent the proliferation nutrient and survival nutrient concentrations, respectively (Mallet and de Pillis, 2006). The cell species' abundance are given by H for the host cells, T for the tumor cells, and I for the immune cells. D refers to the diffusion coefficient of the respective nutrient species indicated by their subscript (Mallet and de Pillis, 2006). The rate constants k indicate the respective consumption rates for each of the nutrient for each of their assigned cells (H, T, and I). The nutrients can also represent activators, inhibitors, or other protein complexes (e.g., enzymes, epigenetic modulators, drugs, chemical exposures/carcinogens, therapies/perturbations, etc.) as chosen appropriate for the model system of interest.

Further, CA such as Conway's Game of Life can stochastically simulate cancer cell kinetics and their multi-scale tumor population dynamics (Poleszczuk and Enderling, 2014). CA are thus tools for computational systems oncology, to monitor tumor growth dynamics under drug perturbations or targeted therapies, in software space (Metzcar et al., 2019. For example, Monteagudo and Santos (2014) demonstrated that CA models can simulate the behavioral dynamics of cancer stem cells (CSCs), which as discussed, are believed to be in large part, responsible for the emergent adaptive behaviors in tumor ecosystems. To further illustrate, in another set of studies, stochastic CA models well-captured the dynamics of avascular tumors under chemotherapy and immunotherapy perturbations and provided computational insights into how drug delivery should be optimized to inhibit tumor proliferation (Pourhasanzade and Sabzpoushan, 2019).

To further illustrate, in a model by Qi et al. (1993), a two-dimensional lattice was used with four discrete states, one denoted cancer cells, one represented normal healthy cell, and the other two represented immune cells interacting with the tumor and host cell environments. Probabilistic rules with non-local and non-homogeneous transition rules updated synchronously, resulted in the emergence of cancer cell behaviors which closely matched experimentally observed Gompertz growth models. Similarly, Kansal et al. (2000) simulated a brain tumor growth model via a three-dimensional Voronoi network, with three discrete states representing three types of malignant cells: proliferating cells, quiescent cells, and necrotic cells. Similar local transition rules and conditions as the model by Qi et al. were then used to model the tumor growth dynamics. The pattern dynamics closely matched those observed in experimental brain tumor data.

A rich repertoire of experimentally validated work on cellular automata-based approaches in tumor modelling has been performed using the Cellular Potts model (CPM). The CPM, also known as the Glazier-Graner-Hogeweg model is a time-discrete Markov chain spatial lattice model for studying complex cellular dynamics in biological populations (Balter et al., 2007). Some pertinent examples of such complex cellular processes include cell-interactions mediated collective behaviors (e.g., collective cell migration, cell fate decision-making/differentiation dynamics, etc.), and multiscale pattern formation systems including cancer morphogenesis and tumor invasion dynamics (Scianna and Preziosi, 2012; Szabó et al., 2013; Hirashima et al., 2017; Rens and Edelstein-Keshet, 2019). The individual cells are represented by simply-connected domains on nodes for a given cell index. The CPM dynamics evolves by updating the lattice configuration one cell at a time based on probabilistic transition rules following a modified Hamiltonian-dependent Monte Carlo simulation/Metropolis algorithm (Balter et al., 2007). The experimental works of Sen's and Bhat's groups have well-supported the applications of the CPM model and similar CA-based computational modelling in decoding the complex multicellular dynamics underlying cancer metastasis and invasive- extracellular matrix (ECM) remodelling (Kumar et al., 2018 [a,b]; Pally et al., 2019). For instance, Pally et al. (2009) validated that the reaction-diffusion mediated multiscale focal adhesion dynamics and ECM-remodelling of breast carcinoma can be accurately model the cancer invasion processes.

Other models of multi-cellular tumor growth systems and tumor angiogenesis (i.e., vascular tumors) have also been successfully reproduced using CA systems (Moreira and Deutsch, 2002). For instance, in a two-dimensional lattice CA, a square topology with a nine-membered Moore neighborhood was used in a tumor growth model by Serra and Villani (2001). The model accurately reproduced in vitro tumor cultures' growth dynamics with varying growth conditions such as the difference after exposure to carcinogens and the resultant development of transformation foci. The model quantified the effects of the chemicals and the change in the culture medium by its exposure to good precision matching those obtained by mean-field theoretic approaches on underlying ordinary differential equations. In principle, these approaches can also be extended to in vivo tumors and patient-derived xenografted tumor modelling for monitoring the responses of targeted precision therapies. These are some of the many examples to illustrate that CA are robust tools in quantitative and computational oncology to model tumor growth dynamics under therapy control and help regulate/adjust clinical decision making towards optimized precision medicine.

**Summary:** Criticality, the state of being poised between regularity and chaos, was illustrated as a hallmark characteristic of cancer processes including EMT switches, cell fate plasticity dynamics, tumor pattern formation, metastatic invasion, and complex dynamics in computational epigenetics (e.g., chromatin-epigenetic modifications during cell state transitions). Cellular automata (CA) was discussed as a powerful computational modelling approach to investigate these critical dynamics in multiscale cancer systems.

## 8. ENTROPY

Entropy is seen as a measure of uncertainty or disorder in traditional branches of physics such as statistical mechanics and thermodynamics, respectively. For instance, a gas of molecules has a higher entropy than its liquid or solid phases because a greater number of rearrangements of its microstates (particles) would correspond to the same macrostate (gas). However, in complex systems theory, (Kolmogorov-Sinai) entropy is discussed as an information-theoretic measuring the flow of information across state-space by the trajectories of a dynamical system. Takens (1981) described topological entropy as one of the traditional measures for chaotic-behavior detection in fluid turbulence. The phase-space flows of the system can then be quantified as a transfer of information. The time evolution of the set of orbits originating from all possible initial conditions of the system generates a "flow" in state-space, governed by a set of n first-order differential equations: $\frac{dX_i}{dt} = F_i(x_1, x_2, \dots, x_n)$, where n is the dimensionality of the space and X is the state-vector characterizing the trajectory of the dynamical system. If we can assign probabilities $P_i$ to each of the possible outcomes in the bifurcations of the system, we can define the information associated with the outcome as given by the Shannon's entropy: $H = -\sum_i P_i log_2 P_i$. When entropy increases sufficiently high beyond some critical value of the governing parameters, a phase-transition can occur as denoted by the bifurcations of the attractor dynamics (e.g., transition from a fixed-point to a limit cycle or, from an oscillation to chaotic attractor). The dynamical systems analog of Shannon's entropy is formally referred to as the Kolmogorov-Sinai (KS) entropy or metric entropy. A system with positive Lyapunov exponents will show a positive KS entropy (Reichl, 1992). There is also another useful entropy measure for dynamical systems known as topological entropy, a variant of the metric entropy, wherein instead of a probability measure space we use a metric space with a continuous transformation. Their uses may depend on whether we are dealing with ergodic, flow preserving systems.

The onset of phase-transitions can be quantitatively measured using an information production rate given by the entropy rate: dH/dt. In chaotic systems, an increasing (positive) entropy rate is observed. Intuitively, the increased entropy rate can be interpreted as a measure of unpredictability and irreversibility in the information flow of the system. Thus, maximal entropy and positive entropy rates can be used as predictors of the birth of complex attractors) (i.e., strange attractors) in the phase-space of the dynamical system. However, some works have established that entropy is not a robust measure of complexity. The Shannon entropy fails to capture the algorithmic content of a dataset and thereby fails as a measure of graph (network) complexity (Zenil et al., 2017). The KS entropy (rate) can quantify the amount of information (flow) from or within system but whether the information flow is causally related or not cannot be inferred. Regardless, they can be used as cross-validation techniques for chaos detection and may be useful for Waddington landscape reconstruction (i.e., quantify cell state attractors using metric entropy measures).

**Summary:** Entropy was introduced as an information-theoretic tool for dissecting cancer cybernetics and complex dynamics in cancer differentiation processes. Topological entropy was discussed as a measure of chaoticity in complex attractor reconstruction.

## 9. SIMULATIONS AND COMPUTATIONAL DYNAMICS

Due to the limited availability of three-dimensional time-series datasets, the study of complex dynamics within cellular (cancer) cybernetics heavily depends on computational simulations. Computational simulations are emerging as powerful tools for reconstructing chaotic attractors and inferring chaotic or critical dynamics in biological networks. As discussed, Sarkar et al. (2019) used simulations of differential equations to infer critical dynamics in simplified cancer-EMT networks. We also discussed the appearance of chaotic oscillations in cancer-immune competitive growth dynamics when time-delay was introduced as a control parameter in simple modelling differential equations (Khajanchi et al., 2018). The emergence of chaotic attractors was suggested as indicators of long-term cancer relapse and the emergence of aggressive cancer phenotypes (Khajanchi et al., 2018). Let us consider an example from the works by Jensen et al. on the use of simulations to detect intracellular chaos in protein oscillations. Their works are an extension of the Goodwin oscillator model (Goodwin, 1965) to cancer-relevant protein systems.

A Transcription Factor (TF) is a protein which binds to the enhancer or promoter regions of a gene of interest with some affinity and forms a complex with RNA polymerase to transcribe the gene. The control of transcription regulates gene expression and its resultant protein translations. Many cancer-related TFs exhibits oscillatory dynamics within cells (Jensen et al., 2012). For instance, the oscillations of the tumor suppressor p53, Wnt, and NF-kB are TFs central to regulating immune response, apoptosis tumorigenesis, and cancer cell division. The works by Jensen et al. (2012) have shown that oscillatory external stimuli might induce chaos and phase (mode)-locking inside cells when coupled to their internal protein oscillations. Using microfluidic cell cultures, Heltberg et al. (2016) delivered periodic

TNF simulation to fibroblasts and recorded the NF-kB nuclear localization by live cell fluorescence imaging. CellProfiler and MATLAB peak analysis algorithms were used to track cells and quantify the NF-kB translocation, where the activation was quantified as mean nuclear fluorescence intensity normalized by mean cytoplasm intensity (Heltberg et al., 2016). The phase locking transitions in an oscillatory manner were observed even amidst noise fluctuations at critical bifurcation points. When the oscillations entered a chaotic state, counter-intuitively, the NF-kB protein was shown to be most effective at activating downstream genes and optimizing their signaling cascades (Heltberg et al., 2019).

The emergence of intracellular chaotic behavior, at the level of protein oscillations, remains highly controversial and subjected to debate. However, the computational simulations paired with experimental models, as performed by Jensen et al. (2012) are the first set of studies showing *deterministic chaos* can drive a nonlinear internal oscillator within cells such as NF-kB with a periodic external signal such as the cytokine TNF. With low level amplitude oscillations of the external driving signal, it can entrain or synchronize with the nonlinear oscillator as indicated by the Arnold tongues observed in its bifurcation diagram (Heltberg et al., 2019). Arnold tongues are regions of parameter space where the NF-kB oscillations are entrained to the external TNF oscillation. Entrainment implies frequency and phase-locking. Outside the Arnold tongue, there is no synchronization. As the TNF amplitude increases beyond a critical threshold, chaotic dynamics can occur as indicated by period-doubling bifurcations and the overlapping of Arnold tongues (Heltberg et al., 2019). The model shown by Heltberg et al. consisted of a negative feedback loop system (with inhibitor IkB$\alpha$) in a single nonlinear oscillator (Heltberg et al., 2019). These simulations show that the strong coupling of two nonlinear oscillators with a negative feedback loop can give rise to complex dynamics in cell states. These findings suggest that a negative feedback loop in protein oscillatory networks may be a sufficient condition for driving chaos in cells.

Lastly, we will discuss simulations and computational modelling in epigenetics as an example of multiscale dynamics in computational/systems medicine. *Computational epigenetics* is a field at its infancy in comparison to simulations of cancer cell population dynamics or patterns of regulatory network dynamics. As discussed, forecasting the long-range interactions in 3D-genome structure, histone interactions, and other epigenetic processes is the key to deciphering cancer stemness networks and phenotypic plasticity in cancer cell fate dynamics and commitments. These emergent behavioral patterns are the drivers of adaptive features in cancer ecosystems such as therapy resistance and intratumoral heterogeneity. One of the best examples of epigenetic control and regulation in tumor transcriptional dynamics is the polycomb memory system in pediatric high-grade glioma.

Current approaches to modelling histone mark spreading dynamics and chromatin looping dynamics include ordinary differential equations (ODEs) modelling the catalytic kinetics with their (experimentally confirmed) rate constants, or stochastic kinetic models., the latter of which remains the most widely employed approach due to the analytical constraints of ODEs. We can consider a simple epigenetic feedback circuit like the antagonistic feedback between H3K27 and H3K36 methylation (Alabert et al., 2020), or a much simpler single histone modification’s methylation or acetylation dynamics for such ODE models followed by some iterative differential equation solver to approximate the solutions (e.g., Euler methods, Runge-Kutta, etc.) (Chory et al., 2019). In the case of more complex, scalable models, like the antagonistic H3K27me2/3 and H3K36me2 circuit, ODE simulation toolkits such as AMICI (combines SUNDIALS and SuiteSparse) and PESTO are available for ODE solving and gradient-based parameter estimation (Alabert et al., 2020). To illustrate, the methylation dynamics of the histone mark H3K79me0/1/2/3 by the enzyme DOT1L (known for impaired functions in leukemias), can be given by the following system of ODEs:

$$\frac{d[me0]}{dt} = -k_{on}[me0] + k_{off}[me1]$$
$$\frac{d[me1]}{dt} = k_{on}[me0] - k_{off}[me1] - k_{on}[me1] + k_{off}[me2]$$
$$\frac{d[me2]}{dt} = k_{on}[me1] - k_{off}[me2] - k_{on}[me2] + k_{off}[me3]$$
$$\frac{d[me3]}{dt} = k_{on}[me2] - k_{off}[me3]$$

Where t denoted time, $k_{on}$ is the forward methylation reaction rate and $k_{off}$ is the reverse reaction rate (also accounts for cell division, nucleosome turnover, and demethylation) (Chory et al., 2019). The brackets [ ] denote the concentration of the specific histone methylation marks, where 0 is unmodified H3K79 and 3 refers to the trimethylation.

In contrast, the more popular set of approaches in epigenetic modelling of chromatin or histone state dynamics involve stochastic models such as Monte Carlo simulations, Langevin dynamics/Random walks, and coarse-grained molecular dynamics. For instance, a recent study has shown that stochastic computational simulations can well predict PRC2 dynamics in glioma systems, even under the presence of antagonistic H3K36 modifications and H3K9me3 marks propagation dynamics (Harutyunyan et al., 2020). The stochastic simulation STOPHIM adopts a bi-modal random walk model of PRC2-mediated histone methylation dynamics across a simulated genomic region represented as a 1D-vector. The study found that H3K27me 2/3 marks which are widely deposited by PRC2 across broad genomic regions, show globally inhibited methylation distribution patterns in H3K27M glioma cells (Harutyunyan et al., 2019; 2020). Although, the model includes cooperativity in PRC2 dynamics, and the simulated kinetics/catalysis of the methylation rates agree with experimental data, chromatin phase-separation and 3D-genomic structural organization is lacking due to the adopted 1D- linear model. The integration of ChIP-Seq and Hi-C data is required to model the 3D conformation dynamics, which is an essential step to forecast critical dynamics (phase-transitions) in histone marks or chromatin states. The next mission for AlphaFold-like algorithms should be chromatin folding and inferring transcriptional states/dynamics from chromatin/epigenetic states

There are other stochastic/probabilistic simulation approaches available in modelling epigenetic states and histone mark spreading dynamics in cancer systems. Some examples include the use of coarse-grained molecular dynamics from chemical master equations over 1D- lattice models with mean-field approaches (for analytic solutions) used for sirtuin-2 mediated acetylation dynamics in simple yeast systems (David-Rus et al., 2009) and Markov Chain Monte-Carlo algorithms (Jost and Vaillant, 2018). The general approach is that a master equation describes the time-evolution of the probability distribution P for times between DNA replication, at which point histone components are distributed to the daughter DNA molecules in conjunction with semi-conservative replication (i.e., half retention of epigenetic marks). The models often assume a two-state or three-state epigenetic marks, where the histone sites are A (Acetylated), U (unmodified), or M (Methylated). At each time-step a random lattice site representing a nucleosome is chosen and one of the biochemical (kinetic) reactions underlying the chemical master equation (CME) are randomly simulated with a probability proportional to their respective rate constants for the methylation or acetylation dynamics (Dodd et al., 2007; David-Rus et al., 2009). Mean-field approximations must be employed to derive equilibria points (i.e., stable epigenetic states/marks). With cooperativity in the epigenetic marks, bistability (presence of two stable equilibria) is observed in the system representing on/off epigenetic states (Lovkvist and Howard, 2021).

As such, algorithmic information dynamics can be employed in prospective studies to treat these epigenetic switching systems as discrete dynamical systems suited for Ising spin-glasses, cellular automata, or artificial neural networks.

If we assume *s*, the density of marked nucleosomes, is always large and exhibits faster dynamics than the unmodified states, one can take to the limit of large number of nucleosomes the Fokker-Planck Equation (FPE) for the probability P. Also, for the general situation where recruitments of enzymes by active or inactive marks are asymmetric, the steady-state distribution of FPE has at most three fixed points on the bifurcation diagram with a cusp-like catastrophe indicative of critical dynamics (phase-transition) (Jost et al., 2014). Further, the FPE approximations have been shown to closely match simulations of the CME approach by Gillespie algorithm. For example, the Gillespie algorithm has been used to model polycomb memory systems in simpler model systems (Lövkvist et al., 2021b). Further details of CME and FPE are provided in the respective citations of the studies, and in Uthamacumaran (2020).

However, what happens with 3D-conformation dynamics and long-range interactions? Multi-protein complexes and transcriptional marks are involved in the dynamics of epigenetic states, where many types of histone co-modifications are involved. Current models are thus simplified to at most two or three histone modifications. Long-range interactions need to be accounted for in the modelling by integrating Hi-C data with ChIP-Seq tracks. For instance, few groups have previously investigated how long-range looping interactions may be involved in H3K9me3 domain (constitutive heterochromatin) formation by use of Monte Carlo simulations coupling nucleosome turnover with methylation kinetics (Chory et al., 2019; Ancona et al., 2020). The stochastic models were able to well-reproduce the chromatin marks seen in experimental ChIP-Seq profiles. Similar studies were shown to reproduce the SETD2-catalyzed H3K36me3 marks from ChIP-Seq tracks, as well (Cermakova et al., 2019). Recent studies have shown stochastic chemical kinetics polymer models using Langevin dynamics to better capture histone methylation dynamics than these above-listed methods (Katava et al., 2021). Regardless of these computational approaches, the lack of 3D-chromatin modelling and a lack of time-points in conjunction with histone spreading dynamics remains the central problem in current computational epigenetics. Predicting critical dynamics in epigenetic remodelling of chromatin states and their resultant cell fate dynamics with limited time-points is a great burden to complex dynamics discovery in (cancer) epigenetic systems.

**Summary:** Various computational models in stochastic dynamics/simulations were discussed for investigating multiscale cancer dynamics, including population dynamics and growth/invasion processes. These coarse-grained stochastic models include molecular kinetics, differential equations-based model systems, Monte Carlo approaches, and Gillespie algorithm. A summary of molecular dynamics in the emerging field of cancer computational epigenetics was also introduced. These simulations can be paired with experimental techniques and other discussed computational toolkits within this paper to better elucidate the behavioral patterns in cancer ecosystems.

## 10. MACHINE LEARNING-DRIVEN CAUSAL INFERENCE

While the above-discussed traditional chaos detection methods can verify if the state-space attractor reconstructed from the time-delay embedding of cancer signals is chaotic, they are bound to dimensionality limits. What happens if multiple chaotic attractors coexist in the signaling/expression state-space? Imagine the computational complexity of dissecting the time-series of a network of thousands of genes or proteins within thousands of cells at once. Identifying chaotic attractors in the state-space of such complex networks is an NP-hard problem. While traditional approaches may fail to dissect these complex networks, model-driven and physics-driven artificial intelligence may provide a solution for causal inference. Recurrent Neural Networks (RNNs) are recently emerging as the state-of-the-art machine learning algorithms for the spatiotemporal prediction of chaotic dynamics and attractor reconstruction in complex time-series datasets. They allow the model-free inference of chaotic dynamics from complex datasets. For example, Reservoir Computing (RC), a type of RNN has recently demonstrated applicability in the Lyapunov exponents prediction of spatio-temporally chaotic systems, such as the forecasting of the KS (Kuramoto-Sivashinsky) equation up to a few multiples of the Lyapunov-time (Pathak et al., 2017; 2018).

RC computing is a merging line between Liquid-State Machines and Echo-state networks, two types of random recurrent neural networks (RNNs). Liquid State Machines (LSM) are a type of spiking neural networks composed of artificial neurons with threshold activation functions (Maass et al., 2002). Each neuron is also an accumulating memory cell of random interconnections. On the other hand, ESNs are random, large, fixed recurrent neural networks. Each neuron within this reservoir network produces a nonlinear response signal. ESNs are equivalent to LSMs from a dynamical point of view. Both parallel approaches were recombined to RC computing, the current state-of-the-art machine learning to predict chaos in time-series (Verstraeten et al., 2007). The RC neural network consists of three distinct layer types: the input layer, the Reservoir, and the Output layer. The Reservoir is a network of nonlinear units forming recurrent loops with random configuration. Only the output layer is optimized by training (adjusting the weights). No Backpropagation is needed for training thus, it is simple and quick (Pathak et al., 2018). Different output layers can be trained for different tasks (i.e., parallel computing).

For a simple reservoir update, consider the input U(n), the states of the reservoir at time X(n), and the output at a given time is y(n). Let W be the connectivity of the nodes of the reservoir. Using some nonlinear function, we can recursively update the network from its data points in the current state. The reservoir update is described by the generic rule: $x(n) = f\left(Wx(n-1) + W^{in}u(n)\right)$ and the network's output computation is given by $y(n) = W^{out}x(n)$. Applications of RC include dynamic pattern classification, chaotic time-series generation, and chaos forecasting (prediction). The Lyapunov exponents and chaotic attractors of spatiotemporally chaotic systems can be attained using RC computing. For example, Pathak et al. (2018) exploited the RC reservoir dynamics to find the Lyapunov exponents of high dimensional dynamical systems, from which chaotic attractors could be reconstructed. Local and global metrics such as the Kullback-Leibler divergence, cross-validation measures, and mean-square error can assess how accurately the chaotic attractor was mapped by the neural network or how well the Lyapunov exponents were predicted for the chaotic system.

In some ways, one can think of predicting cell fate transcriptional dynamics or signalling dynamics as reminiscent of weather forecasting in tumor ecosystems. Both are multi-dimensional patterning fluid systems with multi-scale dynamics. As such, it may be useful, in general, to adopt AI-driven computational fluid dynamics (CFD) and fluid turbulence modelling approaches in the study of cancer patterning/cybernetics. For instance, Ling et al. (2016) used custom Deep Learning architectures with Galilean invariance to approximate the Reynolds' stress tensor in Navier-Stokes Equations flows in turbulent regimes. More recent examples of this includes machine-learned super-resolution analysis and reconstruction of complex turbulent flow fields (Fukami et al., 2019). Fukami et al. used convolutional neural networks (CNN) and a hybrid down-sampled skip-connection/multi-scale (DSC/MS) model to forecast complex fluid patterns. Another example would be the shallow decoder network by Erichson et al. (2020). In such approaches, we can take a few measurements or coarse-grained resolution measurements of the flow dynamics for training the neural networks, and in result forecast/predict its high-resolution flow patterns. However, there are limitations since this is an image-based training method and large, high quality image datasets are required. Furthermore, there are Lyapunov times, windows of predictability, to consider given the 3D-flow evolution of complex structures such as fractal hierarchical patterns and vortices. The reconstruction becomes poorer as we go farther away from the training interpolation region.

More recently, a class of RNNs referred to as liquid neural networks, or liquid time-constant networks, are also emerging as continuous-time neural networks for data-driven time-series forecasting of complex dynamics (i.e., causal inference) (Hasani et al., 2020). These methods remain unexplored in cancer research and modelling/forecasting cancer signaling dynamics. Thereby we should extend these computational models and AI-driven simulation techniques to study cellular patterning systems and chemical turbulence (i.e., intermittent, or spatiotemporally chaotic intracellular flows in morphogens and protein oscillations). There are many other neural networks such as Generative Recurrent Neural Networks with reinforcement learning and other Deep Learning frameworks which can also be trained to detect chaotic attractors in cancer signaling/expression dynamics. The reinforcement learning model is most applicable if the amount of time-resolved data available is very little, wherein the neural network will generate new data which mimics the experimental data for pattern recognition. However, such methods will not be discussed herein.

**Summary:** Various causal inference algorithms and computational systems in the field of machine learning/artificial intelligence were reviewed in this section for capturing causal patterns/relationships in cancer dynamics. These machine intelligence tools include certain types of neural networks such as reservoir computing, liquid neural networks, and recurrent neural networks. These tools should be exploited in pattern discovery in various cancer processes such as decoding cellular dynamics in gene expression state-space (differentiation dynamics), reconstructing protein signaling networks, and deciphering histone/epigenetic modifications in cancer chromatin-state transitions.

## 11. ALGORITHMIC COMPLEXITY

Algorithmic Information Dynamics (AID) is an artificial intelligence platform for causality inference in dynamical systems. AID demonstrates that the algorithmic information of complex networks can be used to steer and reprogram their complex dynamics in phase-space (Zenil et al., 2019; Zenil et al., 2020). AID provides a set of tools to approximate the Algorithmic (Kolmogorov) complexity of these complex networks and control them via merging algorithmic information theory with perturbation analysis in software space. Perturbation analysis can be as simple as the removal of an edge or node from a complex network. A graph network can be represented by a set string or array of binary code. The algorithmic information content of this string/array can then be described by classical measures such as Shannon entropy H(s) or Kolmogorov complexity K(s). The K-complexity, K(s), also known as Kolmogorov or algorithmic complexity quantifies the shortest bits of a string or computer program required to describe a dataset. K-complexity is a robust measure of a network's complexity vastly unutilized in current approaches to network biology (Zenil et al., 2016).

K(s) may be seen as analogous to Shannon entropy as a measure of complexity (or the lack of complexity, i.e., randomness) (Zenil et al., 2019). However, K(s) is a more robust tool than Shannon entropy to measure the complex dynamics of networks. Unlike our current statistical approaches in inferring complex networks dynamics (such as Shannon's entropy or correlation metrics), K-complexity provides causal inference of network topology and dynamics. By perturbation analysis using AID tools, one can identify the sub-structures of complex networks driving their information flow and regulating their topology (and in consequence, the cellular states/phenotypes) (Zenil et al., 2019). Although Shannon entropy can quantify the amount of information in a complex system (network), it does not tell us how causally connected they are. Further, entropy provides no insights into the algorithmic content of a graph network. However, the algorithmic information content of a complex network distinguishes a process as a cause or randomness (Zenil et al., 2019). Furthermore, K(s) does not depend on a choice of probability distribution like Shannon entropy does. Therefore, it is more robust than Shannon entropy in measuring the complexity of a graph networks, such as cancer plasticity networks. Further, we have shown that Shannon's information entropy rates is closely matched to lossless compression algorithms in comparison to algorithmic complexity. K(s) is also emerging as a machine intelligence platform to reconstruct attractor landscapes such as the Waddington epigenetic landscape of biological networks and causal discovery in their network state-space dynamics (Zenil et al., 2019; Zenil et al., 2020).

Formally, the Kolmogorov complexity of a discrete dynamical system s is $K(s) = \min \{|p| : U(p,e) = s\}$, where p is the program that produces s and halts running on an optimal reference universal Turing machine U with input e. K(s) is the length of the shortest description of the generating mechanism (of the network or system). For example, a graph network or system is defined as random (or not having a causal generating program) if the K(s) is about the same length of s itself (in bits). However, *K(s) is semi-uncomputable* and must be approximated using tools from AID.

K(s) can be seen as analogous to a measure of the compressibility or irreducibility of an object such as a string or network, or a dynamical system. Then, K(s) of a network matrix s is the length of the shortest compressed file producing s when decompressing it. Compression algorithms like LZ77, LZ78, Hauffman coding, and LZW (Lempel-Ziv-Welch) are some examples of lossless compression algorithms (Zenil et al., 2019). They are closer to Shannon entropy (rate) estimations than the graph complexity since they can detect statistical regularities within the information system. However, currently no compression algorithm can estimate the K(s) of a complex network since they are not sensitive enough for small perturbations. As such, Block Decomposition Method (BDM) (Zenil et al. 2018) can be justified as the most appropriate method to study graph and network complexity perturbation analysis (Zenil et al. 2019) also providing a more sensitive and robust alternative to limitations of entropy-based statistical compression algorithms such as the LZ and LZW family of compression algorithms.

The Coding theorem method (CTM) is based upon, or motivated by, algorithmic probability and is able to provide an estimation to K(s). However, CTM is computationally expensive (i.e., applicable only to short string or small object sizes). Therefore, BDM is available as an extension of CTM. It approximates the K(s) of a dataset, providing local estimates of the algorithmic complexity (Zenil et al., 2019). Let U be an optimal reference universal Turing machine and p be a program that produces s running on U, then, the Solomonoff-Levin algorithmic probability is given by:

$$m(s) = \sum_{p:U(p)=s} 1/2^{|p|} < 1$$

Then, the shortest program p, K(s), is related to the algorithmic probability by the CTM, which states: $K(s) = -log_2\big(m(s)\big) + O(1)$ (Solomonoff, 1964; Kolmogorov, 1965; Levin, 1974; Chaitin, 1976).

There is also Bennett's logical depth, a measure based on Kolmogorov complexity, defined as follows:

$$Depth_s(x) := min_p\{T(p): l(p) - K(s) \leq s, (U(p) = x)\}$$

While the K-complexity measures the length of the minimal program required to generate the string or graph s, the logical depth measures the fastest program(s)/computation time T, i.e., shortest running time length, needed to generate the system (Bennett, 1988). This is a very interesting measure because it would capture objects in the chaotic regime and place them as having deep structure even when, to some purposes, are random-looking. They are neither the simplest by their emergent behavior nor algorithmic randomness, but their dynamics require computational time to emerge at a usually small critical interval (Zenil et al., 2012).

The set of graph eigenvalues of the adjacency matrix is called the spectrum of the graph. The Laplacian matrix of a graph is also sometimes referred to as the graph's spectrum. Eigenvalues of evolving networks can be computed, and one can observe the graph complexity K(G), where G is the graph representing the string s, versus the complexity of the eigenvalues, to obtain information about the amount and kind of information stored in each eigenvalue (Zenil et al., 2015; Zenil et al., 2018). Further, by assessing the maximum entropy per row of the Laplacian matrix, the eigenvalue which best characterizes the evolving network can be identified. Graph spectral analysis provides a quantitative tool for characterizing attractor dynamics in complex networks. CTM studies dynamical systems in software space, characterizing the effects of perturbations and natural or artificial changes to a system in terms of the changes in the set of the underlying explanatory computational models able to explain the system before and after the intervention (Zenil et al, 2020).

The Block Decomposition Method (BDM) allows to combine the power of statistical information theory and algorithmic complexity hence extending the range of CTM to characterize local but longer-range algorithmic patterns. BDM is defined as: $BDM = \sum_{i=1}^{n} K(block_i) + log_2(|block_i|)$, where the block size must be specified for the n-number of blocks. When the block sizes are higher, better approximations of the K-complexity are generally obtained. Although all methods listed here are applicable to causal discovery in dynamical systems, BDM is the most useful (and robust) tool in AID so far to study K(G) on all objects, including applications of perturbation analysis to graphs and networks, and as such has the potential to provide a computational framework to quantify the causal structure and complex dynamics of cancer networks (Hernandez et al, 2018; Zenil et al., 2019).

To apply BDM on cancer networks/datasets that operates at a discrete and binary alphabet, one can binarize the underlying adjacency matrices with a moving threshold to obtain a vector of networks and associated BDM values. Then apply the BDM on the vectors and obtain a vector of BDM values to work with. By sampling through all thresholds, the process is immune from the arbitrary choice. The identification of the essential features of complex networks- motifs, cliques, and subgraphs, is an NP-complete problem (Zenil et al., 2019). Therefore, identifying cancer stemness networks is NP-complete, in principle, in traditional approaches unless some *a priori* assumptions are made on the underlying data distribution. However, *algorithmic complexity* and the AID toolbox of measures to approximate K(G) avoid such assumptions of predicted data distributions to fit the complex system and find computable candidate mechanistic models. AID provides a robust platform to identify *causal structures* such as chaotic attractors in cancer networks (Zenil et al., 2019).

Perhaps the most elegant aspect of AID is that it provides a computational description of biological information processing. Unlike our traditional perspective of evolution by natural selection, AID provides a view of adaptive processes as *algorithms* steered by causal information dynamics. In systems science, one often uses the term *cybernetics* to denote the study of information processing (dynamics), regulation, feedback, communication, and control in complex systems. As such, cyberneticians often refer to complex systems as control systems, regulatory systems, or feedback systems. Cells, genes, and proteins, the essential structures of information processing in biological cybernetics, can then be treated as computers, programs and codes forming complex multi-scaled feedback loops and hierarchical structures. AID allows causal discovery in such complex systems. Further, in a recent study, AID measures such as BDM have been demonstrated as powerful tools which could highlight evolutionary paths in biological systems. The algorithmic probability reduces the space of all possible mutations and AID was shown able to detect biological pathways more likely to generate mutations versus those which are more stable (Hernandez-Orozco et al., 2018). These findings demonstrate that evolutionary dynamics may be treated analogous to evolving programs in software space. Therefore, AID measures may provide a robust platform to study the cybernetics (information flow) of driver mutation networks and stemness networks in cancer evolutionary dynamics. These approaches should be extended to the study of cancer systems and CSCs to map their cell fate choices and help identify the minimal set of mutations or driver signals required to confer cancer stemness.

**Summary:** In this last section of computational/mathematical approaches, the powerful framework of algorithmic information dynamics was introduced. Algorithmic complexity provides a robust screening tool for cancer dynamics under a computational systems framework, whereas cellular processes can be viewed as programs and cells as computers. Network perturbation analysis using algorithmic complexity measures was discussed as a statistically strong method to identify causal biomarkers governing cancer cell fate dynamics.

## 12. CONCLUSION

In summary, various algorithms for the detection of chaotic attractors in the signaling state-space of cancer networks have been discussed. The basic insights into chaos, fractals, and complex systems have been sowed in the context of cancer dynamics. Although chaos exhibits apparent randomness, it has distinct properties and patterns which distinguish it from stochasticity. More precisely, chaotic systems exhibit emergent structures in their state-space with a (multi)fractal dimension: *strange attractors*. Although mathematical and computational models of cancer dynamics have demonstrated the existence of chaos and strange attractors within cancer cells, their experimental confirmation remains limited. The lack of time-series cancer datasets (largely in part due to technological barriers) and a lack of complexity science in cancer research are fundamental barriers in experimentally detecting complex dynamics in cancer cells. However, there are various emerging ways to acquire time-sequential cancer datasets in single-cell transcriptomics and proteomics, as discussed in the introduction.

A blueprint (tree-diagram) of causal inference in time-series cancer datasets is provided in Figure 2. The general road map to detecting a chaotic attractor (if it exists) in cancer signaling dynamics is such that first the time-traces of the signal of interest such as gene expression from time-resolved single-cell transcriptomics or protein oscillations from live-cell imaging is acquired (Figure 2). Then, it must be embedded via time-delay coordinate embedding to be visualized in a three-dimensional space. Following, various discussed algorithms such as fractal dimension, Lyapunov exponents, and entropy measures, can be applied to verify if the embedded pattern is a chaotic attractor(s). There are other chaos detection tools which were not discussed here and could be useful in dissecting biological cybernetics. One good example would be the 0-1 test proposed by Gottwald and Melbourne (2016). However, given the dimensionality limits of such traditional techniques like time-delay embedding and topological entropy, the review strongly suggests the exploitation of machine learning algorithms like RC networks and liquid neural networks, and artificial intelligence platforms like algorithmic information dynamics (AID) for causal pattern discovery in cancer systems (Figure 2). The techniques outlined in the tree-diagram have widespread applications in systems medicine, including other single-cell multiomics datasets (e.g., protein abundance matrix from CyTOF or histone mass spectrometry, single-cell chromatin modifications matrix from EpiTOF, etc.) (Cheung et al., 2018). Table 2 in the Appendix provides a simplified summary of how the complex systems techniques/tools may apply to different types of datasets and the format of the dataset required for their application.

Chaotic behavior in population dynamics/cellular ecosystems has been predicted as a signature for generating heterogeneity, tumor aggressiveness, metastatic invasion, recurrence/relapse, and therapy resistance (Itik and Banks, 2010; Khajanchi et al., 2018). Further, intracellular chaos has been suggested as a hallmark of cancer progression and aggressivity herein. Jensen et al. demonstrated the flows of protein densities may form strange attractors within cells (Jensen et al., 2012; Heltberg et al., 2016; Heltberg et al., 2019). In extension of their findings, chaos is suggested as a causal mechanism by which tumor phenotypes can acquire adaptive properties and increase their fitness in harsh fluctuant environments. The detection of chaos within cellular oscillations and protein flows are predicted to be indicators of complex dynamics driving cancer networks. Further, chaotic dynamics in a single transcription factor were shown to orchestrate phenotypic heterogeneity and the enhancement of downstream gene signaling (Jensen et al., 2012; Heltberg et al., 2019). Then, the emergence of intracellular chaotic dynamics at the level of protein flows and gene regulatory networks may allow cancer cells to become highly robust to perturbations, conferring adaptive advantages to dynamic environments (resilience), promote their phenotypic plasticity, and generate aggressive phenotypes with therapy resistance.

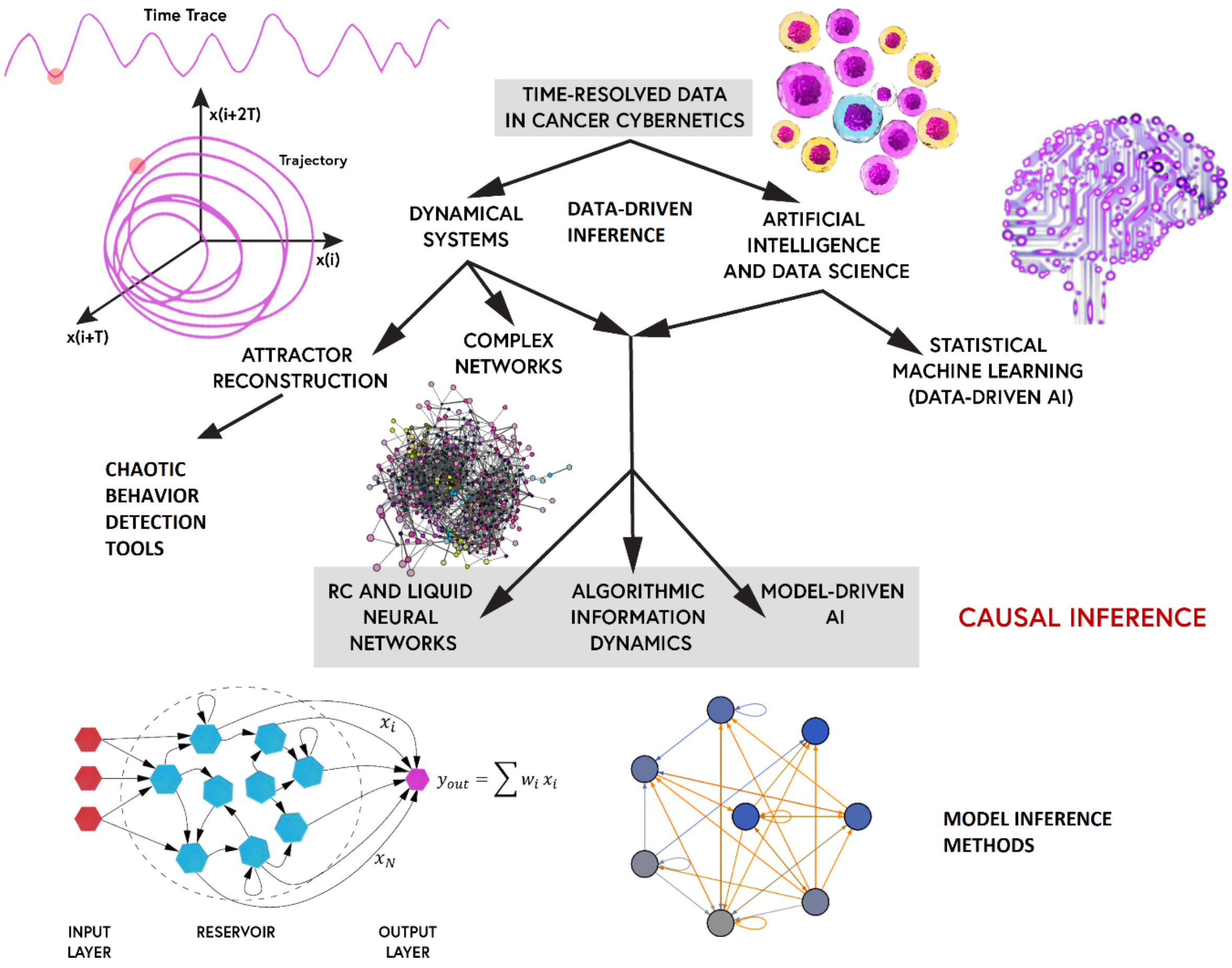

**FIGURE 2 BIOLOGICAL INVERSE PROBLEM.** *The workflow summarizes a blueprint of causal inference methods and measures discussed in the review for systems oncology. Given time-resolved cancer data (e.g., live-cell imaging of protein flows, time-sequential transcriptomic profiling, etc.), we can employ complex systems tools such as dynamical systems modelling or statistical machine learning algorithms for pattern discovery. Dynamical systems approaches include attractor embedding followed by chaotic behavior detection tools as discussed, or complex networks inference. Chaotic behavior detection tools comprises of many approaches discussed in the paper including attractor embedding, fractal analysis, frequency spectra, and Lyapunov exponents. However, these approaches may have dimensionality limits and hence, AI-driven causal inference algorithms are proposed as promising tools for causal pattern discovery in single-cell time-sequential analyses, which include algorithmic information dynamics (i.e., measuring the algorithmic complexity of complex graph networks via perturbation analysis in software space), recurrent neural networks (e.g., RC networks, liquid neural networks, etc.), and model-driven AI (e.g., turbulence modelling/multiscale computational fluid dynamics).*

Therefore, chaotic, or complex dynamics are not suggested as signatures of cancer pathogenesis herein, for which, as well-established, genetic instabilities and epigenetic abnormalities provide a better causal mechanism. Rather, complex dynamics are suggested as signatures of tumor progression and aggressivity in cancer cell fate dynamics. The detection of a chaotic attractor in cancer signaling implies the presence of cellular (disease) state, which is complex, adaptive, and difficult to treat. Hence, perturbation analysis by means of targeted therapies or cellular reprogramming methods can be used to determine which therapy/perturbation results in the loss in instability and complexity of the strange attractor, and thus, help identify effective precision therapies against aggressive cancers like GBM. Network medicine and complex systems analysis provides another tool to help identify these targeted therapies or gene/protein drug targets for the perturbation analysis. If we see the strange attractor (complex dynamics) reduce to more stable attractors such as equilibrium points as indicated by the loss of a fractal dimension of the attractor or non-positive Lyapunov exponents, we may conclude the perturbation is a robust anti-cancer therapy. In principle, such approaches provide a causal framework to not only screen for precision therapies but also control/predict cancer cell fate dynamics and reprogram their phenotypes towards benignity. Further, causal inference methods should be applied to single-cell multiomics and multimodal profiling methods.

The future of mathematical and computational methods for cancer research holds great promise with the emergence of interdisciplinary fields such as computational oncology and systems medicine. The various tools discussed herein illustrate that they provide quantitative insights into complex cancer processes hindering therapy response and contributing to disease progression/aggressivity, including cancer differentiation dynamics, phenotypic plasticity/stemness, and adaptive heterogeneity. Such complex behavioral patterns allow cancers to develop adaptive traits such as therapy resistance, recurrence/relapse, and metastatic invasion leading to cancer-related deaths. The mathematical and computational tools allow clinician-researchers to dissect the complex networks underlying these behavioral patterns and help elucidate putative therapeutic targets specific to these behaviors. Network medicine and attractor reconstruction allow the control and regulation of patient-derived cancer systems both in silico and in experimental settings, to find novel effective strategies to prevent disease progression or permit cancer cell fate reprogramming. Further, causal inference tools such as algorithmic information dynamics, allow clinician-researchers to decode the causal relationships in driver networks steering the multiscale dynamics of cancer ecosystems, examples of which include transcriptional networks controlling cell fate plasticity and stemness discussed in the paper. Such methods allow treatments tailored towards dynamical responses in cancer therapy since they treat cancers as complex dynamic and adaptive diseases, and hence allow time-dependent progression control-predictability of cancer evolution. These approaches is most beneficial to clinical oncology as it would help pave more effective, cancer network-targeted treatments, precision diagnostics and prognosis with longitudinal screening of patients (e.g., blood-sera biomarkers), and thereby provide extension of cancer patient survival rates. We also predict the complex systems framework of computational/systems oncology may help reprogram cancer (stem) cells to benignity.

**ACKNOWLEDGEMENTS:** We thank Dr. Jacek Majewski of McGill University, for the knowledge he granted on glioma epigenetics and computational epigenetic modelling. Thanks to Rik Bhattacharja (Concordia University) for redesigning the figures drafted by AU. Figure 1A was adapted from https://ha0ye.github.io/rEDM/articles/rEDM.html

**AUTHOR CONTRIBUTIONS:** The article was written by AU under the supervision of HZ.

**DECLARATIONS OF INTERESTS:** The authors declare no competing interests.
**DATA AVAILABILITY STATEMENT:** Not applicable (N/A)

## APPENDIX

## COMPLEX SYSTEMS: PATTERNS AND BEHAVIORS

The paradigm of complex systems, a multidisciplinary intersection of dynamical systems theory, data science, and various branches of systems science, remains relatively new to most conventional thinkers. As such, we provide here some rich insights into the general features and properties of complex systems. Complexity science is the quantitative study of patterns, processes, and behaviors in complex dynamical systems.

The term *complexity* can also be interpreted as a measure of computational complexity- how difficult it is to solve a problem using the time, space, and algorithmic resources of a computer. Such a computational problem can be classified as P (polynomial-time) vs. NP (nonpolynomial deterministic), or undecidable (Sipser, 1997). In a branch of complexity theory, known as Algorithmic Information Theory, *complexity* is defined by the length of the shortest description (in bits or qubits) of a system. We refer to this information theoretic as the algorithmic complexity or Kolmogorov complexity (Sipser, 1997). The complexity of a state or system is the minimum size of a program that can generate or describe the system (Sipser, 1997). In the study of cancer and cancer signaling, being interested in first succinct, smallest principles, we are interested in the mechanistic models of gene regulatory network or protein-protein interaction network that confers cancer stemness. As such, algorithmic complexity theory is relevant in science in general. In the context of cancer, a cancer stem cell may be interpreted as the minimum size of a molecular network that confers its stemness properties (i.e., self-renewal, differentiation, and therapy resistance). However, cancer ecosystems comprise of interdependent multicellular networks and not the stem cells alone. Further, they exhibit emergent multiscale dynamics requiring the integration of transcriptomics, proteomics, epigenomics, and longitudinal immune monitoring, making such a task (i.e., to define the algorithmic complexity of the system) an intractable, and ultimately, uncomputable problem. This does not mean, however, that algorithmic complexity theory is unable to provide important insights, and that methods should be discarded *a priori*, or that the interaction of information theory with areas such as dynamical systems should be avoided. Signals detected by classical and algorithmic information theory in the form of criticality markers, may be hallmarks of cancer cell fate dynamics and epigenetic memory systems (i.e., phenotypic switching), respectively, and detecting these hallmarks within the cancer networks may provide a solution to overcome the vast search space of this intractable problem.

Throughout the paper, *complexity* will be primarily referring to either *complex systems* (i.e., nonlinear systems exhibiting emergent/collective behavioral patterns) and/or *algorithmic complexity*, a robust measure of non-randomness and computational irreducibility (Sipser, 1997; Wolfram 2002; Zenil et al., 2022). A system is then defined as a *complex system* (non-random) when its algorithmic complexity is not lower than the length of the shortest system (in bits) able to generate it. Complexity thus provides a framework to quantify and characterize complex processes and multiscale behaviors as information dynamics. *Complexity theory* emerged in the 1980s as a discipline of computer science, while the birth of complexity science or complex systems theory was observed around same time. We now know that algorithms can simulate complex systems, and complex systems can simulate algorithms. There is a fundamental principle of correspondence between algorithms and naturally observed complex systems/processes. The key principle of complexity theory is that certain *algorithms are complex systems* (Sipser, 1997; Wolfram, 1988). As such, both (computational) complexity theory and complex systems theory can be merged in time to a single systems framework, which shall be referred to as complexity science herein (Wolfram, 1988). The central paradigm of complexity science is the realization that any complex system can be represented as *information* dynamics (Zenil, 2022).

The study of chaotic dynamics and its emergent patterns is at the heart of complexity science. In 1963, Edward Lorenz, Ellen Fetter, and Margaret Hamilton pioneered modern chaos theory by studying the unpredictability of hydrodynamic flows in simplified Rayleigh-Benard convection systems as an approximation to weather turbulence forecasting (Lorenz, 1963). There have been earlier pioneers such as Henri Poincare (the father of dynamical systems theory), George D. Birkhoff (ergodic theorem), Mary Cartwright, John E. Littlewood, and Aleksandr Andronov (theory of oscillations and bifurcations), amidst others, who have seen the *fine structures of chaos* emerge before them when studying certain classes of nonlinear differential equations. These *fine structures* are today known as *strange attractors*. The works of Kolmogorov, Sinai, Arnold, Hénon, Smale, Anosov, and others further expanded the study of chaotic dynamics/attractors into the domain of differential geometry (topology) and complex systems such as fluid turbulence. The iconic picture of chaotic behavior and that of a strange attractor is the Lorenz attractor, a multifractal structure observed in the phase-space of their simplified set of differential equations. The Lorenz attractor's intuitive description underlies the famous butterfly effect, i.e., the flap of a butterfly's wings in Brazil may result in the birth of a tornado in Texas.

Most complex systems are high-dimensional *chaotic systems*, or rather, a high-dimensional chaotic system is usually considered a *complex system*. Chaotic systems are inarguably the most complex (difficult to describe) of *complex systems* but not the most complex in the algorithmic randomness sense but the most interesting as they are algorithmic simple, but their unfolding appears complex and sometimes even random (Zenil et al., 2012). The reason being that chaotic systems are unpredictable in the long-term and finding solutions to their underlying equations (if they exists) are analytically intractable, necessitating heuristics or computational methods. For instance, the Lorenz attractor is a computationally solved approximate solution to its differential equations (at certain critical parameters). Notions such as intractability, unpredictability, irregularity (fractality), aperiodicity, irreducibility, and undecidability are inherent to complex systems dynamics. Yet *deterministic chaos* presents another feature of most complex systems: *emergence*. Emergent behaviors define simple, nonlinear interactions at a lower scale spontaneously giving rise to complex hierarchical structures or multi-nested (recursive) patterns in higher scales (Wolfram, 2002). Perhaps this is counter-intuitive as apparently random fluctuations and disorder at one scale of interaction builds up to collective organized behaviors at another. For this reason, chaotic behavior is often referred to as an irregular or strange causal order. Scaling dynamics and the self-organization of these multi-scaled, hierarchical structures are observed in emergent systems (i.e., more is different) (Anderson, 1972). Think of phase-transitions, the flocking of starlings (Vicsek model), the stigmergy of ant colonies, the sudden outbreaks of pandemics, social network dynamics, Jupiter's Great Red Spot, and tumor formation; these are emergent behaviors (Czirok and Vicsek, 2000). Note: Some scholars distinguish emergence from self-organization such that emergence denotes the macroscopically observed collective spatial behaviors in a complex system which cannot be deduced from its many interacting microscopic constituents, while self-organization denotes the emergent patterns/collective behaviors in the temporal dimension. Thus, emergence or synergy denotes the *collective behaviors*/dynamics of nonlinearly interacting parts within a dynamical/feedback system.

To illustrate emergence (or self-organization), individually, ants resemble random walkers with very little intelligence. However, collectively, swarms of ants exhibit large-scale coordinated behaviors and super-intelligence. In the case of the starlings, the flight of individual birds may appear random-like and disordered. However, nearest neighbor-interactions allow flocks of birds to undergo a phase-transition and exhibit aggregate patterns and behavioral processes resembling the complex multi-scale dynamics of hydrodynamic flows (Vicsek et al., 1995). In the mathematical study of complex systems dynamics, we are interested in the causal patterns to which emergent behavioral patterns are confined to in state space. The self-organization of strange attractors, like the Lorenz attractor, in state-space of the dynamical system is the classic example. Finding these emergent patterns and structures in a complex dynamical system is analytically intractable. Therefore, complexity science advocates the use of computational algorithms and machine intelligence to help find approximate solutions to such complex systems. Furthermore, from an algorithmic perspective, we can distinguish chaotic behavior from randomness by its Kolmogorov complexity (Sipser, 1997) and we shall explore methods to compute estimates of K-complexity in complex dynamical systems.

## NONLINEAR DYNAMICS AND CHAOTIC ATTRACTORS

The drastic change of behavior in a nonlinear dynamical system by tuning its critical parameter(s) can demonstrate increasing complexity in attractor dynamics. As we change some set of order parameters, first the dynamical system may exhibit solutions which converge from any initial condition to a static equilibrium value(s) (i.e., a fixed-point attractor). As the values of the bifurcation (order) parameter(s) increase further, the fixed-point attractor undergoes bifurcations and transitions to an oscillation (limit cycle, or periodic attractor). As we further increase the critical value of the bifurcation parameters, the oscillation may eventually make a transition into the chaotic regime (Shaw, 1981). We refer to these transitions of an attractor at some critical point of the bifurcations as *symmetry breaking*. Symmetry breaking defines the phenomenon in which small fluctuations acting on a system beyond a critical threshold abruptly changes (decides) the system's fate, by determining which branch of a bifurcation is taken (Shaw, 1981; Thompson and Stewart, 2002). A stable attractor of a nonlinear system can undergo symmetry-breaking when its order parameter(s) exceed a critical point, above which complex dynamics may emerge. The key insight to understand here is that a fixed-point or a periodic attractor can give birth to a strange attractor as it loses stability (control) and bifurcates from its critical points (Grebogi et al., 1987; Gleick, 2008). Chaotic solutions may be indicators of the most complex types of dynamics where the system exhibits irregularity and long-term unpredictability (Thompson and Stewart, 2002; Gleick, 2008). As such, intuitively, if a cell phenotype or cell signal behaves as a chaotic attractor in state-space, it is difficult to control and predict. However, its cell fate dynamics or signaling patterns are confined to the specific set of states bound to the fractal (strange) attractor. That is, although individual trajectories of the system may exponentially diverge apart, the global orbit/structure of these trajectories are bound to a finite set of state-space (basin of attraction), due to the stretching-folding of phase-space into a fractal architecture (Gleick, 2008; Strogatz, 2015).

One way to visualize the birth of attractors is by observing its bifurcations in phase-space as we tune its order parameters. Another approach is to simply consider the system as a network of coupled nonlinear oscillators. Now consider an external oscillation as a control parameter of this network of coupled oscillators (i.e., cells or genes/proteins). An external oscillatory signal (driving force) can be used to create phase-locking and synchronize the oscillators (i.e., collective dynamics) at some lower critical threshold of the external signal's oscillation frequency (Strogatz and Stewart, 1993; Strogatz, 2004). If the frequency of the external oscillator is strongly coupled to the system, and is tuned at its natural frequency, resonances are observed (increasing amplitudes). However, as we keep increasing the external periodic signal above some critical frequency threshold, the system of coupled oscillators will exhibit Arnold tongues that either causes the entrainment to the external periodicity or results in *aperiodic oscillations* which may further bifurcate towards *chaotic behavior* (Jensen et al., 2012;

Heltberg et al., 2019). Spatiotemporal chaotic behavior can be detected in various routes in a complex system, one approach is to observe *period-doubling bifurcations* in its phase portrait. In the context of oscillators, as discussed, a *broad band frequency spectrum* may be a robust signature of chaos. While periodic oscillators exhibit well defined peaks with amplitudes corresponding to their signal intensity (e.g., concentration of proteins in time, or trajectory of a cell fate transition), a chaotic oscillator will show a broad continuous spectrum. In more complex dynamics, anomalous multifractal scaling may be observed in the system's energy (frequency) spectrum (Jensen et al., 2012; Strogatz, 2015). The power law scaling exponent can be extracted from the slope of the line observed in the log-log plot. However, with many interacting oscillators (cells, molecules, genes/proteins, etc.), it may not be easy to observe such features.

## COMPLEX NETWORKS AND MULTISCALE DYNAMICS

Many of the complex emergent behaviors in tumor ecosystems are attributable to a subpopulation of adaptive cells referred to as cancer stem cells (CSCs), found within tumor ecosystems with distinct properties, including self-renewal, therapy resistance, phenotypic plasticity, and the ability to differentiate to multiple heterogeneous phenotypes (Plaks et al., 2015; Xiong et al., 2019). The phenotypic plasticity of CSCs are regulated by their nonlinear interactions with the dynamic tumor microenvironment. Their stemness depend on the signaling networks of their stem cell niche, a complex tumor microenvironment comprised of immune cells, extracellular matrices, blood vessels networks (angiogenesis), and healthy cell networks (Rosen and Jordan, 2009; Plaks et al., 2015). The microenvironmental cues can remodel the three-dimensional chromatin structure within the cellular states by various types of chemical post-translational modifications including histone tail alterations, promoter-enhancer looping, and differential DNA methylation, collectively referred to as *epigenetic modifications*. Epigenetic modifications can be transmitted across cell divisions, and the stability of these epigenetic memory systems govern cellular identity and transcriptional dynamics in tumor ecosystems (Flavahan et al., 2018; Meir et al., 2020). The phenotypic plasticity observed in CSCs and their differentiated heterogeneous phenotypes are governed by epigenetic memory systems. As such, phenotypic plasticity is also referred to as *epigenetic plasticity* (Flavahan et al., 2018). Elucidating the causal mechanisms by which epigenetic switches maintain and promote cancer cell fate dynamics is an active research hotspot for systems medicine.

Due to the advancement of single-cell multi-omics and availability of high throughput epigenetic datasets such as scRNA-Seq, histone mass spectrometry, CyTOF/EpiTOF, scChIP-Seq analyses, WGBS (whole genome bisulfite sequencing/methylome sequencing), scATAC-Seq, and Hi-C chromatin capture, our recent understanding of single-cell cancer epigenetics have revealed that pediatric cancers are molecularly distinct from their adult counterparts due to a greater epigenetic burden (Schwartzentruber et al., 2012; Wu et al., 2012). Perhaps the strongest evidence to the role of epigenetic modulation of cancer cell fate decisions would be pediatric high-grade glioma (pHGG) epigenetics (Schwartzentruber et al., 2012). Some key insights into the epigenetic and molecular underpinnings of pHGGs are provided herein as a model-system of epigenetic complexity and complex adaptive behaviors in tumor ecosystems, and to help elucidate some of the mathematical models and computational techniques discussed in the later sections. Although in complex systems such as cancers one cannot separate transcriptional dynamics or proteomics signaling from epigenetics, we shall briefly explore glioma epigenetics to acquire some intuition to how adaptive chromatin states give rise to transcriptional heterogeneity and tumor plasticity in cancer ecosystems. Further, as will be discussed, unlike pattern discovery in transcriptional dynamics, computational epigenetics is an emerging sub-domain of systems oncology which necessitates a special class of algorithms and toolkits pertaining to the study of *critical* phenomena/behaviors. *Criticality* is the state of being poised between regularity and chaos, marked by sudden phase-transition(s) above some threshold control/order parameter of the system (Note: we say regularity instead of order, as it has been traditionally coined, since chaotic behavior is a type of complex, causal/temporal order).

pHGG such as atypical teratoid rhabdoid tumor (ATRT), embryonal tumors, diffuse intrinsic pontine glioma (DIPG), and glioblastoma (GBM) are invasive lethal diseases of the central nervous system (Mackay et al., 2017). They show distinct recurrent oncohistone mutations in the genes encoding histones H3.1, H3.2, and H3.3 (Schwartzentruber et al., 2012, Wu et al., 2012), with the G34R/V (glycine 34 to arginine or valine) and K27M (lysine 27 to methionine) variants indicative of clinically-relevant pathological subgroups. These two mutant oncohistones show altered posttranslational modifications on two key lysine (K) residues of the H3 tail, K27 and K36 due to amino acid substitutions as indicated above. These histone residues regulate complex cellular processes, including developmental genes and embryonic/stem cell differentiation. Further, the emergence of these histone variants also show age and neuroanatomical dependence (Mackay et al., 2017; Deshmukh et al., 2021) (Figure 3).

Roughly 80% of pHGGs of the CNS midline structures (i.e., including the pons, thalamus, and spine) show H3K27M or rarely to isoleucine, substitutions (K27M/I) (Deshmukh et al., 2021). We also see H3.1/2K27M mutations in acute myeloid leukemia (AML), one of the primary hematological malignancies in children. In contrast, G34V/R oncohistones are specific to histone H3.3 mutation (*H3F3A*) which occurs in about 30-50% of cortical pediatric high-grade gliomas and mainly target the temporoparietal cortex (Deshmukh et al., 2021) (Figure 3). There are H3 wild-type gliomas primarily emerging in the fronto-parietal lobes with H3K27M mutations and/or somatic mutations in isocitrate dehydrogenase 1/2 (IDH1/2), thus resulting in excess metabolic enzyme production of 2-hydroxyglutarate (Deshmukh et al., 2021). These epigenetic circuits also reveal how the glioma metabolome forms a bidirectional feedback system with the H3K27 and H3K36 PTM cross-talk, which regulate transcriptional dynamics by recruitment of distinct readers, writers, and erasers (i.e., chromatin remodelling enzymes and proteins). Most chromatin-modifying enzymes require substrates or cofactors that are derived from metabolic intermediates (Kinnaird et al., 2016). This should be self-evident given that our diet/nutrition and drug intake are the primary control mechanisms of epigenetic circuits. The epimetabolic and epiproteomic rewiring in cancer cells provides evolutionary advantages in major cellular decisions, such as proliferation and cell fate differentiation by modulating nuclear transcription, and as such form an interconnected complex adaptive system.

The antagonistic feedback loop between H3K27 and H3K36 methylation dynamics has been revealed as the central regulator of gliomagenesis and glioma progression/cell fate decision-making (Schwartzentruber et al., 2012; Deshmukh et al., 2021). The primary remodelling enzymes mediating these feedback loops are the impaired polycomb repressive complex 2 (PRC2) (H3K27 trimethyltransferase) and SETD2 (H3K36-specific trimethyltransferase) epigenetic memory systems (Schwartzentruber et al., 2012; Huang et al., 2020). There are also enzymes such as NSD1/2/3 mediating the intermediate methylation states in H3K36 di-methylation. A global loss of H3K27me3 due to the inhibition of PRC2 and in conjunction, an aberrant overexpression of H3K36me2 (mediated by NSD1/2) are epigenetic signatures of pHGGs (Schwartzentruber et al., 2012; Harutyunyan et al., 2020). However, how these epigenetic changes affect higher order chromatin structure and oncoprotein signalling or transcriptional dynamics of cancer stemness genes remains unelucidated. The goal of systems oncology should be to employ complex systems theory and tools from artificial intelligence to infer causal patterns in the epigenetic memory systems (e.g., 3D chromatin conformation and histone mark spreading dynamics) in collective behavioral dynamics/processes such as cancer cell fate decision-making. As will be discussed later, we require computational tools and simulations tailored towards critical dynamics to achieve this complex task. As mentioned, the cellular cybernetics of multicellular adaptive systems like tumors show complex ecosystem dynamics (i.e., collective behaviors). That is, many scales of information dynamics (networks) must be integrated to understand their emergent behaviors and for optimal clinical decision-making in the treatment of these dynamic diseases. Thus, we should integrate the spatiotemporal dynamics at the transcriptomic, metabolic, proteomic, and multi-cellular interactions with these epigenetic datasets to forecast chromatin-state dynamics in complex diseases like cancers

For instance, single-cell molecular profiling and proteogenomics characterization of 218 pediatric brain tumor samples of various histological subtypes including 25 high grade gliomas showed distinct molecular features and signalling patterns in agreement with the transcriptional signatures seen in scRNA-Seq profiles (Petralia et al., 2020). The study also profiled the tumor phosphoproteome, the phosphorylation/kinase activity acting as on/off switches to signalling proteins and observed an upregulation of MEK/ERK/AKT kinases, warranting clinically relevant therapeutic targets. Further, these findings reveal that the information dynamics across different scales of cancer processes, whether it be the transcriptome, epigenome, metabolome, or proteome, to some extent capture interdependent characteristic patterns for causal discovery (Armingol et al., 2021). However, the degree of this correlation depends on the cell type context and microenvironmental complexity. For instance, most proteins expressed in cell communication networks including the cell surface receptors and their ligands are transcribed by the cells, and hence, we can infer protein interaction networks from single-cell transcriptomics (Armingol et al., 2021). We can further dissect the cancer cybernetics at other post-translational modifications at the proteomic and epigenomic levels, including the acetylome, ubiquitylome, methylome, glycoproteome, microbiome, immune interactome, etc. For instance, the role of the gut-immune axis in brain cancers has recently been better elucidated. Further, pHGGs show high amounts of immune cell infiltration, and hijacked immune-inflammatory signals for mediating their tumor microenvironmental dynamics and therapy resistance (immune evasion). There are many other layers of complexity identified in patient-derived liquid biopsies such as the circulating tumor cells, dormant tumor cells (i.e., quiescence), and cancer-mediated exosomes/extracellular vesicles in the maintenance of tumor cybernetics (Li and Nabet, 2019; De Angelis et al., 2019; Park and Nam, 2020).

However, amidst all these multi-scale dynamics in cancer cybernetics, epigenetic datasets require the most attention as its causal mechanisms severely lack understanding. Further, deciphering the epigenetic patterns underlying cancer cell fate dynamics may hold promise to controlling and reprogramming cancer phenotypes and cellular decision-making. Current approaches to cancer therapy including chemotherapy, and radiation therapy are ineffective in the treatment of pHGG. Some recent progress in cancer immunotherapy, and in specific to oncolytic viral therapy are showing potential promises in adult GBM as seen with the PVSRIPO trial (Desjardins et al., 2018). We are also seeing the emergence of epigenetic therapies such as Protein arginine methyltransferase 5 (PRMT5) inhibitors showing promise in disrupting patient-derived GSCs with greater sensitivity on the proneural GBM subtypes (Sachamitr et al., 2021). However, understanding the epigenetic control of GSCs (glioma-derived stem cells) may be the key to preventing glioma progression and reprogramming aggressive cancer cell states to stable phenotypes.

For example, recent in vitro studies have shown the reprogramming of U87MG human glioblastoma cells into terminally differentiated neurons using a small molecule cocktail consisting of forskolin (cAMP agonist), ISX9 (promotes neuronal cell fate differentiation), CHIR99021 (GSK3-inhibitor), I-BET 151 (BET proteins inhibitor), and DAPT ($\gamma$-secretase /Notch inhibitor) (Lee et al., 2018). More recently, Gao et al.

(2019) used a combination of Fasudil, Tranilast, and Temozolomide (FTT) cocktail to reprogram patient-derived GBM cells into neuronal-like phenotypes with induced expression of the same neural-specific TFs (i.e., NGN2, ASCL1, etc.) achieved by the above-discussed small-molecule cocktail. Thus, as suggested by these molecular targets and cell fate behavior reprogramming experiments, there are certain signaling pathways and transcription programs such as the GSK3/Wnt pathway and Notch signaling, controlling epigenetic plasticity and differentiation dynamics in glioma systems. However, with a wide set of drug targets such as those used by Lee et al. (2018), it becomes ambiguous which signals are critical for the cell fate reprogramming. As such, we should employ larger drug screens with similar small-molecules or CRISPR screens to identify more robust combinations. Regardless, these studies suggest that perhaps similar lineage-specific transcription factors or chromatin modifiers can force unstable, cancer stem cells or stalled cancer attractors like the pHGGs towards terminally differentiated stable phenotypes (Figure 3). Epigenetic drug targets such as inhibitors of chromatin modifying proteins and histone modification enzymes, such as bromodomain protein inhibitors and CBX inhibitors are also speculated to serve as drugs capable of alleviating the developmental/differentiation blockade in oncohistone variants of pHGGs (Nagaraja et al., 2019). Further, EZH2, the functional enzymatic component of PRC2, is required for glioma stem cell maintenance (Suvà et al., 2009). As mentioned, polycomb dynamics and polycomb memory systems are dysregulated in the oncohistone variants of pHGGs. A deeper insight to the epigenetic and transcriptional circuits regulating these transcriptional programs in relation to polycomb dynamics (and cellular chromatin structure) could help us reprogram and control cancer ecosystems towards benignity.

While the epigenetic profiles are distinct in pediatric gliomas from adults, it remains unelucidated whether similar histone-histone interactions (i.e., combinatorial histone marks), or epigenetic and transcriptional programs also drive the disease in both age groups. For instance, it has been shown that adult GSCs can converge into an epigenetic state reminiscent of paediatric GBM via selective downregulation of H3.3 expression (Gallo et al., 2015; Lan et al., 2017). Hence, similar complex networks may be driving gliomas (cancers) in children and adults, or perhaps the driver complex network's topology undergoes changes from one group to another. These findings warrant further understanding of the epigenetic drivers of glioma stemness/differentiation networks and epigenetic plasticity in complex adaptive tumors. Complex systems approaches will help us quantify cancer cell fate reprogramming and cell fate determinations/decisions as patterns of complex network dynamics and transitions in the stability of attractors on the networks' state-space (landscape). We should repeat the glioma cell fate reprogramming experiments towards forced neuronal lineage commitment discussed above with pediatric glioma samples, including the oncohistone variants, as they are believed to be stalled in their developmental trajectories and more refractive to differentiation/reprogramming (Deshmukh et al., 2021). These studies should be conducted along with healthy controls, to determine whether any side-effects or targets also affect the healthy cells.

**Summary:** The multiscale dynamics and collective behavioral patterns (attractors) underlying cancer cell fate decisions were introduced herein. The biology-heavy section was intended for non-cancer experts such as physicists, mathematicians, and computational scientists interested in modelling cancer dynamics. Pediatric high-grade gliomas were introduced as a toy model to illustrate the complex multiscale behaviors in tumor cybernetics from epigenetics to proteomics and gene expression patterns. The relationship between 3D-chromatin conformation dynamics, histone/epigenetic modifications, and gene expression profiles in tumor ecosystems was discussed herein. The discussion of key concepts such as epigenetic plasticity, phenotypic switching, and epigenetic memory systems would be pertinent for some of the discussions of mathematical treatments/models in the main paper. These biological insights could help interdisciplinary thinkers investigate causal patterns in multiscale cancer systems/dynamics using our discussed complex systems toolkits/approaches in the following sections. Further, computational epigenetics is at its infancy in cancer research, and the summarized biological discussions herein could help promote its research program.

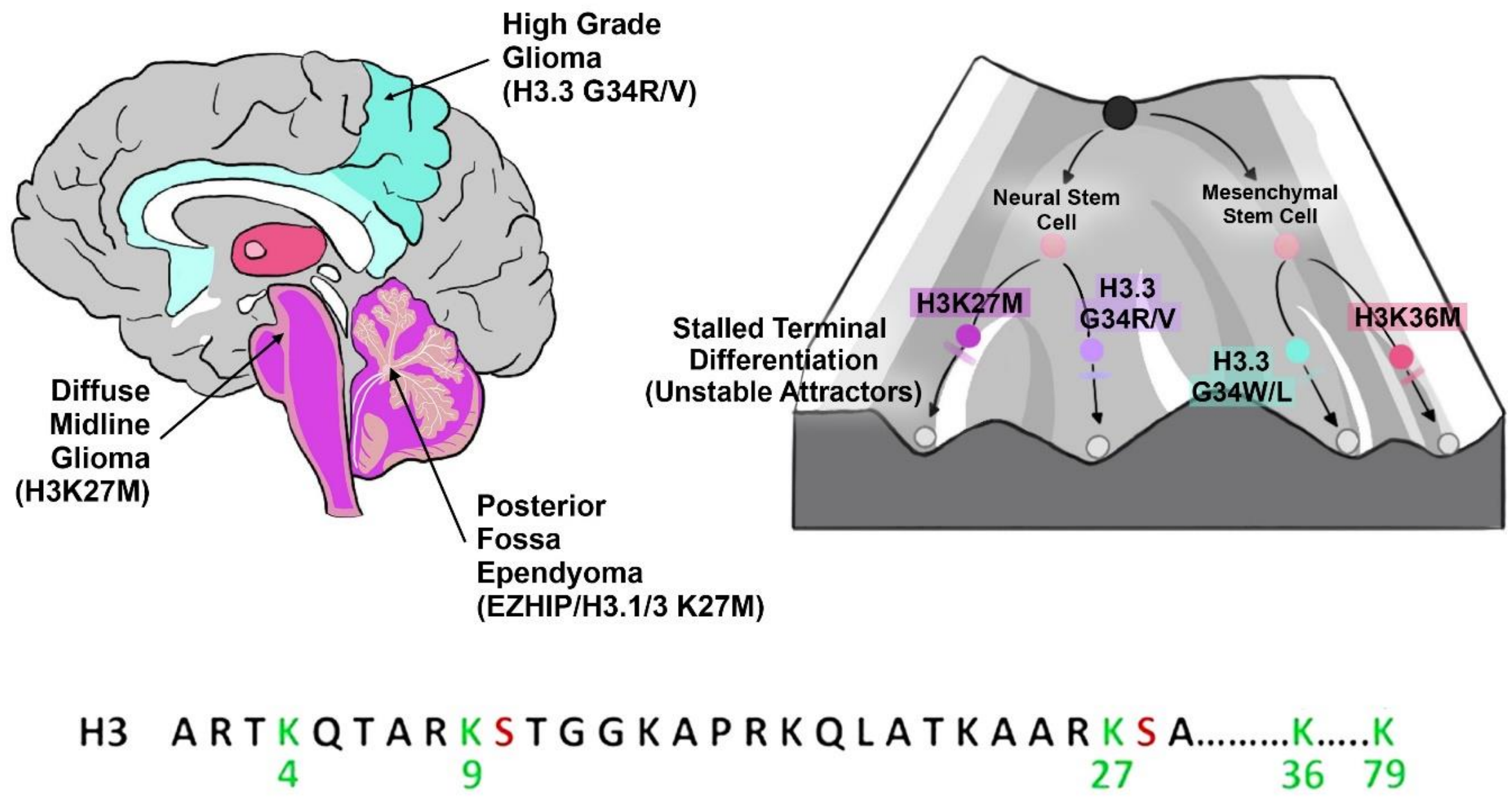

**FIGURE 3. DIFFERENTIATION DYNAMICS IN PEDIATRIC GLIOMA SYSTEMS.** *On the left, a schematic of the discussed epigenetic variants of pediatric high-grade gliomas (pHGGs) are shown with their corresponding brain regions recapitulating altered neurodevelopmental differentiation circuits. The corresponding Waddington landscape for their stalled differentiation dynamics is shown to the right. The cancer cell fates are shown as stalled attractors on the landscape (gene expression or signalling state-space) resembling stem cell states. Below, a string of the amino acid sequence of the histone tail H3 code is provided with the sites of the recurrent epigenetic mutations in these pHGGs. Some of these epigenetic modifications correspond to active chromatin marks with transcriptional activity while others are repressive marks (inhibited gene expression). The polycomb system is an essential regulator of pHGG differentiation dynamics. The toy-model system provides the biological insights underlying the complex dynamics and mathematical concepts discussed in the paper.*

## LYAPUNOV EXPONENTS: INTRACELLULAR PROTEIN FLOWS AND LIVE CELL IMAGING

The Lyapunov exponents can be applied to a wide scale of cancer processes such as the study of protein flows and collective cell (ecological) dynamics. At the scale of protein flows and collective cell migrations, the dynamics behave as a continuum analogous to the hydrodynamic flows of fluids. For instance, when using time-lapse imaging to map cell fate trajectories (as in the case of CSC differentiation mapping), collective cell migration must be considered. Similarly, when imaging the concentrations of fluorescently labelled proteins in a population of CSCs undergoing differentiation (e.g., intracellular morphogenetic flows), the protein patterns within the cells must be treated like a fluid. A vector, velocity field $v(x,t)$ must be defined for such fluid-like systems from the time-lapse imaging data. Various computational tools and software exist for defining the velocity field from imaging data (e.g., CellProfiler). The Lagrangian flow map F $(x_0, t)$ can then be defined using the velocity field of the intracellular protein flow or collective cellular migration (flocking) patterns, as given by:

$$F(x_0, t) = x_0 + \int_{t_0}^{t} v(F(x_0, \tau)d\tau.$$

The Lagrangian flow maps the initial positions $x_0$ of the cells or protein flows at time $t_0$ to their final position at time t (Serra et al., 2020). Using the Lagrangian flow map, we can compute a Lagrangian (continuous) analog of the Lyapunov exponents known as the finite time Lyapunov exponent (FTLE) for the flow trajectories, given by:

$$FTLE(x_0, t) = \frac{1}{T} ln \left\{ \max \left( \frac{|\nabla F(x_0, t)\delta x_0|}{|\delta x_0|} \right) \right\}$$

Where $\nabla F(x_0, t)\delta x_0 = \delta x_t$, T=$t - t_0$, $\nabla$ is the del operator (gradient of F here), and the |·| denotes the Jacobian of the terms inside with respect to initial position. The FTLE measures the maximum trajectory separation rate between the initial position and a neighboring state (protein or cell) starting at $x_0 + \delta x_0$ over time t for continuous dynamical systems (Serra et al., 2020).

**FREQUENCY SPECTRA**

As discussed in the main text, the interaction strength (coupling) between two oscillators is typically increased by tuning the amplitude of an external oscillator (Heltberg et al., 2019). As the interaction strength between the two oscillators increase, highly complex phenomena may emerge. The range of frequencies of interactions widen, and a period-doubling sequence of bifurcations towards chaotic behavior may occur. In general, a broadband frequency spectrum is observed when oscillators exhibit chaotic dynamics (Figure 1B). An ideal example of this is the Kolmogorov energy spectrum for isotropic, homogeneous fluid turbulence. However, complex turbulent flows, in real-life systems, exhibit intermittency and multifractals, as denoted by an anomalous power law scaling. Since irregular dynamics can be observed in the frequency spectra of complex dynamic systems, specific/general features of the frequency (power) spectral analysis is not as robust as the other discussed chaos detection methods in the review. Hence, we have allocated this space in the appendix to discuss some areas of relevance to which frequency analyses may be useful in dissecting cancer dynamics.

To illustrate a biologically pertinent example, recent findings suggest certain protein flows within cells can exhibit *chemical turbulence*. Chemical turbulence defines the emergence of spatiotemporally chaotic patterns observed in protein-mediated reaction-diffusion systems (Rössler, 1977; Kuramoto, 1984; Halatek and Frey, 2018). Individually, the time-traces of proteins may exhibit Brownian motion. However, collectively, certain protein flows may display fluid-like complex hydrodynamics (Rössler, 1977; Kuramoto, 1984; Halatek and Frey, 2018). A broad band energy spectrum was observed in the chemical turbulence regime of protein fluids within cells reminiscent of the Kolmogorov spectrum of isotropic fluid turbulence (Bohr et al., 1998; Halatek and Frey, 2018). These findings were confirmed in both, simulation and in experiment (Denk et al., 2018; Glock et al., 2019). The emergence of chemical turbulence in cell patterning systems further blur the scaling problem in fluid turbulence.

Fourier analyses have shown the absence of regularity in the oscillations of a cancer signal may be an indicator of chaotic dynamics. For example, an attractor reconstruction via Takens's theorem (time-delay embedding) performed on rat prostate culture oscillations confirmed the presence of chaotic behavior with a positive Lyapunov exponent. Cellular micromotion was analyzed using an ECIS (Electric cell-substrate impedance sensing), and the corresponding time-series oscillations' Fourier spectra were shown to be broad and flat due to a lack of dominant harmonics. A lack of periodicity and the emergence of a broad frequency spectrum are thereby considered characteristics of chaotic oscillations (Posadas et al., 1998). Even if the system exhibits complex dynamics, as is the case for complex turbulent flows, multifractality may be detected with an anomalous scaling of the broad-band spectrum. Thus, a simple method for chaotic-behavior detection is to subject the time-series signal traces to an FFT (Fast-Fourier Transform) algorithm such as the Cooley-Tukey algorithm. FFT algorithms are built into MATLAB as a 'fft' command, or virtually found in any programming language or graphical analyses software (See Appendix). The method also allows the extraction of power spectral density analysis, which may be more useful for classifying more than one system exhibiting complex dynamics (i.e., spot characteristic patterns distinguishing the two systems). The efficiency of this method also relies on how much the signal is sampled. The resolution of time sampling, and noise filtering must be considered in the Fourier transform analysis of time-series.

**Summary:** Patterns of frequency spectra in chaotic systems were briefly introduced in the context of cancer signaling dynamics and cellular oscillations. Although a broad-band frequency spectrum may be a signature of chaotic systems, most complex systems exhibit irregular features in their spectra.

**CRITICAL DYNAMICS AND NETWORKS**

Criticality is the governing principle underlying *phase transitions*. For instance, think of the tangible properties (i.e., density, pressure, etc.) of a liquid in comparison to a gas or solid of the same homogeneous substance. These are first-order phase-transitions. A steam of water molecules and an ice cube are two distinct *phases* of the same system. The critical points are the points at which the distinct phases are at equilibrium. There are large density *fluctuations* near the critical point of a system (Sethna, 2006). Under a microscope, there would be patches of liquid embedded into patches of gas molecules when the critical temperature is approached. A practically insightful model to visualize phase transitions and quantify their dynamics are spin glasses such as the Ising model, consisting of a set of coupled, interacting magnetic dipole spins on a lattice, where each spin configuration corresponds to some energy. Below or above a *critical temperature*, symmetry-breaking occurs resulting in a phase-transition from a random spin orientation to alignment (order). However, the optimization (finding the ground state energy) of a 2D Ising model is an NP-hard problem. Think of biological complex systems, where one deals with an order of magnitude in $\gg 10^{23}$ molecules, and hence, the Ising model becomes combinatorially complex (Mezard and Montanari, 2009). However, an effective mean-field theory can be derived for energies below a certain cut-off of the system (critical point) using renormalization groups (Mezard and Montanari, 2009).

Power laws are the signatures of criticality. They correspond to straight lines in log-log plots and are characterized by the generic distribution $n(s) = ks^{-\alpha}$, where k is some constant, n(s) is the size distribution, s is the size (variable of interest) and $\alpha$ is the scaling exponent. Debated evidence to the universality of critical dynamics include earthquakes, solar flares, stock market crashes, social networks, financial/economic networks (i.e., Pareto's principle), pandemics, the internet, etc. as they obey power-law distributions and scale-free networks (Bossomaier and Green, 2000). For example, consider the Pareto's principle where the rich get richer in a power-law behavior. This occurs due to the interconnectivity of socioeconomic networks (i.e., hubs formation). One of the mechanisms for criticality is known as *preferential attachment*, where whether you grow or not in time depends on how big/connected you already are. Another mechanism for critical dynamics is *self-organized criticality*

*(SOC).* SOC defines the tendency of a complex system, often a slowly driven nonequilibrium system, to spontaneously gravitate towards critical phase-transition points for a wide variety of initial conditions (Bossomaier and Green, 2000). Scale-free biological networks tend to exhibit hub markers or hub genes, which are putative therapeutic targets and control switches of cellular dynamics/cybernetics.

To obtain an intuitive picture of SOC, we can consider the Per Bak's sandpile model. Bak et al. (1987) demonstrated using a cellular automaton that each time a grain of sand is added to a pile of sand on a checkerboard/lattice, if the number of sand grains on a square exceeds a critical threshold (tipping point), the pile topples forming an avalanche (in a multidimensional model) and gives sand to its neighbors. The distribution of these avalanche sizes and their lifetimes form *power laws*. As we approach larger and larger numbers of sand grains, in the scale of a million or more, Tibetan mandala-like fractal patterns emerge. Recall that fractals are essentially power law scaling, as well. Therefore, power laws are measures of the tendency of complex systems to *transition to chaos*. A clear example would be the transition of a fluid flow from a laminar phase to a turbulent state. The phase transition to turbulence demonstrates an abrupt change of behavior as some order parameter (the Reynolds number) reaches a critical value and undergoes symmetry-breaking bifurcations. In Kolmogorov's model of isotropic fluid turbulence, the energy spectrum exhibits a power-law decay in the inertial subrange with a critical exponent of -5/3, given by: $E(k) = C\varepsilon^{2/3}k^{-5/3}$, where $\varepsilon$ is the energy flux, k is the wavenumber, E is the energy of the eddies and vortices, and C is some constant (Ruelle, 1995). The turbulent eddies and vortices breakdown into a fractal hierarchy of smaller eddies and vortices within this range (i.e., the Richardson-Kolmogorov energy cascade) (Ruelle, 1995).

Critical dynamics can occur at many scales of cancer cybernetics, from cellular states driven by gene regulatory networks (GRNs) to chromatin states driven by 3D-chromatin organization and epigenetic regulations. In fact, power law scaling, the signature of critical dynamics, suggests scale-invariant patterns and behaviors. Critical dynamics have been observed in complex signaling networks. Most biological networks exhibit non-trivial topologies including critical or scale-free networks, wherein the network degree distribution obeys power law behaviors, where only a few nodes (hub nodes) have numerous edges while the majority have fewer and fewer links (Barabasi and Oltvai, 2004). These topologies are robust to failure when random nodes are added or perturbed, but vulnerable to the failure of the hubs. The network connectivity in many cancer driver gene networks and protein-protein interaction networks (PPIs) are suggested of power law relationships in their degree distribution (Barabasi and Oltvai, 2004). Let P(k) give the probability that a selected node has exactly k links. The degree distribution is obtained by counting the number of nodes N(k) with k-links and dividing by the total number of nodes N., The degree distribution of scale-free networks obeys a power law $P(k){\sim}k^{-\gamma}$, where $\gamma$ is the degree exponent (Barabasi and Oltvai, 2004). Smaller the $\gamma$, the more important the role of the hubs in the network.

In general, universal properties of scale-free networks exist if $\gamma < 3$, where the dispersion of the distribution $\sigma = \langle k^2 \rangle - \langle k \rangle^2$ increases with the number of nodes, and higher the degree of robustness against accidental node failures or random perturbations (Newman, 2003; Barabasi and Oltvai, 2004). There are many other biological network topologies exhibiting power-law degree distributions, such as the Watts-Strogatz small-world networks and the Barabasi-Albert random networks, not discussed herein (Barabasi and Oltvai, 2004). For instance, the Watts-Strogatz small-world networks are observed in metabolic networks and social networks (Barabasi and Oltvai, 2004). Further, there are network centrality measures (e.g., betweenness, eigenvector, closeness, hub-score, etc.) and modularity (community structure detection) available as measures to explore the regulators of information flow and multi-nestedness of complex dynamic networks, respectively. Network centrality measures such as betweenness and eigenvector centrality can be used to identify the regions where these hubs occur in scale-free or complex networks (Freeman, 1979; Landherr et al., 2010).

Critical dynamics have been demonstrated in epigenetic processes, as well. As such, the study of critical systems may be most relevant to the emerging field of computational epigenetics discussed in the main text. Criticality is seen in 3D- chromatin spatial organization which is captured by Hi-C data contact maps and super-resolution imaging (Lieberman-Aiden et al., 2009; Mirny, 2011; Boettiger et al., 2016). The promoter-enhancer regions in chromatin Hi-C contact maps were fitted by power law decay models (Lieberman-Aiden et al., 2009, Mirny 2011, Ron et al., 2017). That is, their contact probability as a function of genomic distance forms a power law decay. We see the emergence of hierarchically organized structures in these enhancer loops called topologically associating domains (TADs) revealing a scale-free pattern (Ron et al., 2017). The domains' boundaries were shown to act as regulatory insulators which control phenotypic transcriptional programs by preventing the expression of outside of the enhancer domain. The TAD borders are also enriched genes with high transcriptional dynamics, as well as cohesin and CTCF binding sites, thus supporting the loop-extrusion model (Fudenberg et al., 2016).

There is an open question of whether histone modifications are also poised at criticality? Computational models reveal that a cusp-like phase transition (catastrophe), as seen in EMT dynamics modelling, can be seen in the simulated bifurcation diagrams of chromatin state dynamics (Jost, 2014). These models capture how the recruitment dynamics of epigenetic modification enzymes allow the formation of long-lived (stable) epigenetic memory systems and the stability of epigenetic/cellular identity. Further, they show that critical dynamics with cusp-like catastrophes (tipping points) in the system’s bifurcation diagrams, analogous to those observed in Ising spin models, defines the epigenetic plasticity of cell fates (Jost, 2014). Epigenetic switching during phenotypic/developmental transitions and patterns of cell fate behaviors (network dynamics), as seen in cancer ecosystems, may thus be explained by criticality and require criticality detection tools. We must further understand how the bistability in epigenetic/chromatin states transitions to the multistability and metastability seen in phenotypic landscapes (transcriptional dynamics). It is proposed herein that the mechanism underlying this stability bifurcations is a *criticality to chaos transition* from epigenetic/chromatin states to transcriptional dynamics, in tumor ecosystems. Dynamical systems theory and complex systems tools are thus required to elucidate how the critical point bifurcations evolve to the complex dynamics seen in disease-state epigenetic landscapes. Furthermore, phase-transition is not to be confused with the liquid-liquid phase separation seen in certain nuclear transcriptional condensates and protein complexes/transcription factors (Zhang et al., 2019). The liquid-liquid phase separation of intrinsically disordered proteins (IDPs) are emerging as therapeutic targets for reprogramming oncogenic cell fates (Shin et al., 2018; Klein et al., 2020).

**CHAOTIC OSCILLATIONS**

In the main text, the works by Jensen et al. were discussed as examples of pairing computational simulations with empirical cellular data to investigate intracellular chaotic oscillations/dynamics. Some fine details of these studies and their prospective suggestions are put forth in this appendix subsection. To find the regions of parameter space that exhibit chaotic flows, Jensen et al. first computed the standard deviation in the NF-kB amplitudes from each time series and found the parameter points at which this grew discontinuously, as the external TNF amplitude was increased. Within these regions, they further tested for chaos by computing the divergence of trajectories that started near the initial points, using deterministic simulations. Parameter regions where such trajectories diverged exponentially were defined as chaotic regions. All deterministic simulations were performed by numerically integrating a set of dynamical equations characterizing the TF-negative feedback loop using the Runge–Kutta fourth-order method, and for optimisation, some of the equations were simulated using Euler integration (Heltberg et al., 2019). All stochastic simulations of NF-kB dynamics were simulated by the Gillespie algorithm. For noise in the external TNF oscillations, Langevin simulations of the different oscillations were used.

These simulations concluded that the flows of protein densities may form strange attractors within cells. Chaotic dynamics in cellular signaling were suggested to allow adaptive heterogeneity to emerge within cell systems. The major findings of these simulations on chaotic dynamics in cellular protein oscillations can be summarized as follows:

- Chaotic dynamics in the oscillation of a single TF NF-kB regulates downstream genes causing their cascading expressions and protein production. The TF plays a vital role in immune homeostasis and cancer stem cell niche dynamics.
- Chaotic dynamics upregulates low affinity genes.
- Chaos increases the efficiency in protein complex formation. The formation of protein complexes are essential for transcription and gene activity regulation.
- Chaos generates dynamical heterogeneous populations of cells and increases their survival in harsh, fluctuant environments. This provides adaptive selection of chaotic cell states to cytotoxic drugs such as chemotherapies.

As suggested by these findings, chaotic dynamics can be used by cancer cells to selectively adapt to their harsh environmental conditions and upregulate certain proteins or specific protein complexes needed for tumor growth and cancer stemness/plasticity. To verify these computational models, these findings should be experimentally validated within cancer cells, especially given that these protein signals are critical driver signals in cancer stemness networks and phenotypic transition/EMT programs. Furthermore, we know of many cancer subset-specific molecular driver network patterns. For instance, Suva et al. identified a core set of four essential transcription factors governing GBM stemness and GSC decision-making (Suvà et al., 2014). The simulations and computational models performed on the NF-kB model can be adapted to these GSC transcription factor networks.

## LIMITATIONS AND PROSPECTS

There are many technological and financial barriers to such time-resolved cancer data acquisition. However, what we propose is the time-series measurements of a single patient-derived biopsy's cell culture by means of multi-modal profiling techniques. For certain techniques like live-cell imaging of protein flows or time-lapse imaging of in vitro cellular differentiation, this could be relatively feasible. However, for techniques such as time-resolved transcriptomic profiling, traditionally it may have been more difficult as the preparation of a new cell culture (even if from the same patient tumor) at a different time point may exhibit different set of dynamics. Regardless, we could always perform in vitro differentiation experiments with cancer cell cultures or organoids subjected to different growth conditions (e.g., in stem cell media versus differentiation media) or drug perturbations, and acquire their single-cell measurements (transcriptomics, proteomics, single-cell epigenetics/chromatin modifications profiling, cell patterning imaging flows, etc.) as distinct time-points. A good example would be the time-sequential bulk RNA-Seq data from murine models used by Rockne et al. (2020) to forecast acute myeloid leukemia (AML) development/progression. Since this was a leukemia model, longitudinal blood samples were easily acquirable from the mice to perform the RNA-Seq.

However, in the case of difficult tumors such as high-grade gliomas and other brain tumors, transcriptomic screening of biopsies at different time points are not feasible with human patients. As such, we could either perform time-sequential measurements on xenograft mice models with human patient-derived tumor biopsies, or alternatively, we can perform in vitro experiments with differing growth media conditions or pharmacological perturbations and/or cell passages to mark the time points, as mentioned. Surgeries can be performed on the xenografted mice tumors over the time-span of days if not months to analyze time-course expression patterns. A recent example of mice tumor models-based time-resolved RNA-Sequencing studies includes the study of BRCA1-associated mammary tumorigenesis dynamics by Bach et al. (2021).

Thus, recent advances are bypassing these technological limitations in scientific causal discovery. There are emerging droplet microfluidics-based techniques such as single-cell metabolically labeled new RNA tagging sequencing (scNT-seq), proposed to offer massively parallel time-resolved single-cell analyses within the same cell samples (Qiu et al., 2020). Another emerging technique is the combined use of fluorescent-reporters and live-cell imaging to acquire time-series profiles of selected genes (or protein) expression dynamics with sampling time rates in the hours scales. For instance, Krenning et al. (2021) developed a method combining live-cell microscopy and FACS-analysis of the FUCCI fluorescent reporter system with scRNA-Seq to acquire high-resolution, time-resolved transcriptomic profiles of mitotic cells at the M-G1 phase-transition. Similar methods can be exploited to acquire high-resolution time-resolved transcriptomics for other types of cellular processes and phenotypic transitions observed in cancer cell fate dynamics. These droplet microfluidics-based techniques and fluorescent-reporters based live-cell imaging techniques should be exploited in the time-resolved differentiation mapping of cell fate trajectories and their time-resolved single-cell multiomics from tumor biopsies/cancer stem cells in the conditions described above (Qiu et al., 2020).

To conclude, chaos/complex dynamics are suggested as hallmarks of cancer stemness and cancer progression and should be screened in the gene expression/protein signaling state-space of cancer stemness networks. Moreover, *criticality* has been proposed as a causal mechanism driving cancer epigenetics and chromatin state organization. The limits of current simulations approaches have also been presented in forecasting the dynamics of epigenetic memory systems in cancer cell fate decisions and differentiation dynamics. Further, although the oncohistone profiles of pediatric gliomas are molecularly distinct from those of adult patients, whether the same driver pathways are steering their cell fate decisions remains an open query. As opposed to our current dogmatic approaches based on snapshot statistical patterns and correlations, the detection of underlying causal attractors would pave a novel research programme, with targeted dynamical therapies in the frontier of systems oncology and precision medicine. Elucidating the complex networks and attractor dynamics governing the transition from criticality (in epigenetic regulation) to chaos (in transcriptional/cell fate dynamics) in CSCs may pave the reprogramming of cancer cell fates to benignity.

## RELATIONSHIP BETWEEN DISCUSSED TOOLS AND METHODS

The various chaos detection tools and methods for causal pattern discovery are interrelated. For instance, entropy is related to fractal dimension as a measure of chaoticity. Higher topological entropy implies higher uncertainty and information flow (Note: thermodynamics uses the term disorder. However, in complexity and statistical physics, the term irregularity or uncertainty/unpredictability is preferred to denote high entropy since chaos is a form of causally complex order). Similarly, a higher fractal index implies higher statistical self-similarity but also higher irregularity. Therefore, entropy and the fractal index are both measures of irregularity in complex dynamics. Topological entropy is also related to the Lyapunov exponents of a chaotic system. As discussed, a positive Lyapunov exponent is a signature of chaotic/complex dynamics. The Margulis-Ruelle inequality and Pesin's entropy formula show that the entropy of a measure that is invariant under a dynamical system is obtained by the total asymptotic expansion rate of the system. The exponential expansion rate of the phase-space dynamics of a chaotic system (i.e., the stretching-folding of state-space) is measured by the Lyapunov exponent(s). The topological entropy is bound by the sum (integral) of positive Lyapunov exponents for a chaotic system (assuming ergodicity) (Hasselblatt and Pesin, 2008). The Takens' embedding theorem or CCM are not directly related to these chaos discovery tools, rather it is the fundamental step taken in the study of dynamical systems to reconstruct the underlying attractor (state-space), given a time-series signal (i.e., attractor reconstruction) (Sauer, 2006). The time-embedding of the signal is then subjected to these various chaos detection tools such as fractal index, Lyapunov exponents, entropy, etc. or causal inference methods such as algorithmic complexity measures or machine learning algorithms for pattern discovery. Takens' theorem and CCM have dimensionality limits, hence, machine learning algorithms, such as the neural networks or physics-model driven methods discussed in the main text are emerging as attractor reconstruction methods for complex dynamical systems.

Algorithmic complexity, K(G) for a graph network G, remains the most robust causal inference tool in the study of computational/complex systems. In algorithmic information dynamics, entropy rate is analogous to lossless compression algorithms, while algorithmic complexity estimates by the Block Decomposition Method (BDM) provide a more robust, and accurate quantification of complex dynamics (Zenil et al., 2019). Further, we discussed various machine learning tools for causal pattern discovery such as liquid neural networks and reservoir computing. It should be noted that these tools are tailored towards finding statistical associations/patterns and not necessarily causal inference. As such, algorithmic complexity remains the central algorithm discussed in the paper optimized for causal inference in complex systems/networks. There are other tools such as the methods developed by Judea Pearl (e.g., Bayesian networks) and certain types of deep learning architectures that could fall within causal inference methods not discussed in the review (Pearl, 2009).

**TABLE 1: GLOSSARY**

| METHOD | DESCRIPTION |
|---|---|
| Takens' theorem | A technique for embedding the time-series signal in state-space using a time-delay in one of its coordinates. Convergent Cross Mapping is an embedding algorithm implementing Takens' theorem, applicable on complex networks. The technique has dimensionality limits and hence, should only be limited to a few signals with predicted chaotic dynamics. |
| Denoising Algorithms | Any algorithm intended for noise reduction. Can range from filtering and preprocessing tools (interpolation, smoothening, etc.) to wavelet-analysis methods. Imputation algorithms are emerging as popular candidates. Not discussed in detail since it consists of a wide range of algorithms, the applicability of which depends on the type of dataset and system of interest. |
| Lyapunov Exponents | Measures how fast two initially close points on a chaotic trajectory exponentially diverge apart in time. Positive Lyapunov exponent(s) are characteristic signatures of chaos. |
| Fractal Dimension | Fractals are the geometry of chaos. A fractal is a geometric pattern exhibiting statistical self-similarity (i.e., power law scaling) across many length and time scales with a fractional (non-integer) dimension. It is used as a measure of irregularity, roughness, and complexity. Some algorithms to estimate the Fractal Dimension include the Box-counting method, Fourier analysis-based approaches, and the sandbox method. |
| Multifractal Analysis | If more than one fractal dimension is required to describe the complexity of the system, multifractal analysis is required. These approaches are most applicable for time-series analysis. The local Holder exponents and the Hurst index are pertinent measures. Wavelet Transform-based methods remain the most popular tools for identifying these multifractal statistics. |
| Fast-Fourier Transform (FFT) | The frequency and power spectra of time-series signals can be acquired using FFT. The FFT algorithm decomposes a time-series into its constituent frequencies. Chaotic systems generally exhibit a broad frequency spectrum. |
| Criticality | Power laws are indicators of critical dynamics, a state of hierarchical self-organization poised between regularity and chaos. When certain complex systems surpass their critical point, they gravitate towards chaotic dynamics. The Ising model is discussed as a powerful tool to model criticality in cancer gene expression and patterns of network dynamics. |
| Entropy | Maximal entropy and a positive entropy rate are observed in dynamical systems exhibiting increased chaotic flows in phase-space. They could be indicators of phase-transitions to chaotic dynamics and/or the birth of complex attractors. However, entropy is not a robust measure of network (graph) complexity and may fail to distinguish randomness from chaoticity. |
| Computational Modelling and Simulations | The pairing of simulations/computational modelling with data science is the central principle of complexity science. Herein stochastic simulations such as the Monte Carlo methods and Gillespie algorithm were discussed for simulating chemical kinetics and molecular dynamics. |
| Recurrent Neural Networks (RNN) | Reservoir Computing (RC) networks and liquid neural networks are the state-of-the-art Deep Learning Networks for time-series forecasting and spatiotemporal prediction of chaotic dynamics from complex, multidimensional datasets. |
| Algorithmic Complexity | Also known as the Kolmogorov complexity (K(s)), is a measure of the length of the shortest description of a dataset (e.g., a string, an array, a network, or dynamical system) or the shortest program needed to generate the dataset. Various algorithms exist for estimating the K-complexity. CTM and BDM (Block Decomposition Method) are alternatives to statistical compression algorithms and are native to n-dimensional complexity. |

**TABLE 2: DATASET FORMAT FOR COMPLEX SYSTEMS METHODOLOGIES**

| TECHNIQUE/METHOD | TYPE OF DATA | NUMBER OF OBSERVATIONS | LONGITUDINAL OR DISCRETE-TIME | NUMBER OF PARAMETERS |
|---|---|---|---|---|
| **Takens's theorem/Convergent Cross Mapping** | Individual | Rich | Both | Minimum 1 dimension for discrete-time and 3 dimensions for longitudinal; and time-delay parameter |
| **Lyapunov Exponents** | Individual or Mean | Rich | Longitudinal | 1-2 parameters (dynamical variable and time) |
| **Fractal Analysis** | Individual | Scarce or Rich | Both (mainly Discrete) | 2 for Box counting technique |
| **Fast-Fourier Transform** | Individual or Mean | Scarce of Rich | Both | Minimum 2 dimensions (time and variable of interest) |
| **Entropy** | Individual or Mean | Scarce of Rich | Both | 1 or more; a priori assumption of statistical distribution for Shannon entropy |
| **Ising Model/Spin Glass** | Mean | Scarce of Rich | Discrete | 1 or more; mean-field approach/a priori assumption of statistical distribution |
| **Cellular Automata (CA)** | Individual | Scarce of Rich | Discrete | 1 or more |
| **Recurrent Neural Networks** | Individual | Rich | Both | Minimum 2 (time and dynamical variable) |
| **Stochastic Simulations** | Individual or Mean | Scarce | Discrete | Statistical Distributions (a priori assumed) |
| **Differential Equations** | Individual or Mean | Scarce | Longitudinal | 2 or more (time and variables); discretization or assumptions are required for analytical solutions |
| **Block Decomposition Method** | Individual | Scarce of Rich | Discrete | 1 or more |
| **Algorithmic Perturbation Analysis (Graph Network Complexity)** | Individual or Mean | Scarce of Rich | Discrete | 1 or more |

The table summarizes the type of data (individual counts or mean), the number of observations (rich or scarce), continuous or discrete-time, and the parameter estimations required for the major complex systems tools and chaos detection measures discussed in the review. The table is a generalization as variations may apply to context-dependence and different systems or processes of interest. In general, these tools can be applied for the attractor reconstruction and network analysis of single-cell cancer multiomics datasets. Some datasets such as histone mass spectrometry or single-cell RNA-Seq require log-normalization of abundance/expression scores, and hence there is pre-processing of the data structure/matrix required. The number of parameters will also vary from one dataset to another depending on the variables of interest. While numerous network inference methods exist ranging from discrete-time networks (e.g., Boolean networks) to continuous-time/longitudinal network analysis, only graph network complexity is listed for illustration purposes. Further, the number of observations is subjected to debate as rich is a qualitative term with system-dependence (e.g., in single-cell datasets, rich implies hundreds if not thousands of cells per sample). Further, it is mentioned if the technique requires a priori assumptions such as fitting the dataset/observations to some statistical distribution.

## SOURCE CODE FOR DETECTION TOOLS:

### 1) CODING AND BLOCK DECOMPOSITION METHOD:

The algorithmic complexity calculator: BDM can be computed on binarized and normalized adjacency matrices (e.g., gene expression matrices) or binary arrays:

http://complexitycalculator.com/

https://www.algorithmicdynamics.net/software.html

https://pybdm-docs.readthedocs.io/en/latest/

### 2) CONVERGENT CROSS MAPPING (TIME-DELAY EMBEDDING):

https://mran.microsoft.com/snapshot/2018-06-22/web/packages/rEDM/vignettes/rEDM-tutorial.html (Sugihara et al.) (rEDM package and tutorial for time-delay embedding)

### 3) LYAPUNOV EXPONENTS:

https://pypi.org/project/nolds/

https://www.mathworks.com/help/predmaint/ref/lyapunovexponent.html

https://blog.abhranil.net/2014/07/22/calculating-the-lyapunov-exponent-of-a-time-series-with-python-code/

4) **FRACTAL DIMENSION (BOX-COUNTING ALGORITHM):** (step by step guide and code for FD calculation)

Frederic Moisy (2021). boxcount (https://www.mathworks.com/matlabcentral/fileexchange/13063-boxcount), MATLAB Central File Exchange.

### 5) RC COMPUTING:

https://github.com/pvlachas/RNN-RC-Chaos

The public version in the above provided GitHub supports forecasting of multi-dimensional time-series.

### 6) FAST-FOURIER TRANSFORM:

https://towardsdatascience.com/fast-fourier-transform-937926e591cb

Bulti-in MATLAB FFT function, use fft (x), for some data matrix x.

### 7) MULTIFRACTAL ANALYSIS:

https://www.mathworks.com/matlabcentral/fileexchange/39069-hurst-exponent-estimation